\documentclass[10pt,twocolumn]{article}
\usepackage{graphicx}

\def\gee{ \, \lower 1mm\hbox{$\,{\buildrel > \over{\scriptstyle\scriptstyle\sim} }\displaystyle \,$}}
\def\lee{ \, \lower 1mm\hbox{$\,{\buildrel < \over{\scriptstyle\scriptstyle\sim} }\displaystyle \,$}}
\def\|{\partial}

\def\Oo {\displaystyle}
\def\varkappa {{\scriptstyle\partial}\! e}

\begin{document}

\twocolumn[

\begin{center}
 {\Large Proceedings of Conference ``\textit{\underline{Dynamics and evolution of disc
galaxies}}'',

\medskip

Pushchino, Moscow region, Russia, May 31 -- June 04, 2010}
\end{center}\medskip

 \centerline{\Large \bf Dynamics of galactic disks in the nonaxisymmetric dark halo }

 \medskip

\begin{center}
 {\bf  Alexander Khoperskov$^\dagger$ $^\ddagger$, Mikhail Eremin$^\dagger$,
 Serge Khoperskov$^\dagger$,

 Marija Butenko$^\dagger$, Sergej Khrapov$^\dagger$}
\end{center}

 %\centerline

\medskip

 \centerline{\bf $^\dagger$ \textit{Volgograd State University, Russia} }

\medskip

 \centerline{\bf $\ddagger$ e-mail: \ khoperskov@volsu.ru}

\medskip

\centerline{ABSTRACT}
\medskip
 {
\textsc{The results of numerical simulations of the dynamics of
galactic disks, which are submerged into nonaxisymmetric dark
massive halo are discussed. Galaxy disks dynamics in the
nonaxisymmetric (triaxial) dark halo were investigated in detail
by the high resolution numerical hydrodynamical methods (TVD \&
SPH) and N-body models. The long-lived two arms spiral structure
generates for a wide range of parameters. The spiral structure is
global and number of turns can be 2-3 in depends of model
parameters. Morphology and kinematics of spiral structures were
investigated in depends of the halo and the disk parameters. The
spiral structure rotates slowly and the angular velocity varies is
quasiperiodic.}
 }

 \bigskip
 ]

\begin{center}
  \textbf{1. Dark massive halo and spiral structure }
\end{center}

The realistic models of disk galaxies require a massive enough
halo. There are the results of numerous estimations of halo mass
using various physical approaches
[\ref{Bottema-Gerritsen-1997!NGC6503-model},
\ref{Khoperskov-etal-2003}, \ref{Mikhailova-etal-2001!z-str}]. It
is essential that the galaxy mass halo and mass of the disk are
comparable and frequently, halo mass even exceeds mass of the disk
inside of a sphere of optical radius $R_{opt}$ ($R_{opt}\simeq
(3-6)\cdot r_d$, $r_d$ --- radial exponental scale of stellar disk
[\ref{Zasov-etal-2004}, \ref{Khoperskov-Tyurina-2003!MW},
\ref{Khoperskov-2002!Late-type},
\ref{Khoperskov-etal-2010!z-str}]).

The absence of central symmetry in distribution of mass and
potential is a feature of some galactic dark halo
[\ref{Keller-etal-2008}, \ref{Newberg-Yanny-2005},
\ref{Newberg-Yanny-2006}, \ref{Newberg-etal-2007},
\ref{Law-Majewski-2010}, \ref{Reshetnikov-Sotnikova-2000},
\ref{Gnedin-etal-2005}, \ref{Brown-etal-2008},
\ref{Olling-Merrifield-2000}]. Generally we may consider triaxial
halo, oblate or prolate halo [\ref{Bullock-2002},
\ref{Kuhlen-etal-2007}].

Spiral structure formation in galaxies is important tasks of their
physics. There are numerous mechanisms of spiral structure
formation such as central stellar bar
[\ref{Patsis-etal-2009!dynamics-barred-spiral}], swing
amplification mechanism [\ref{Bottema-2003!Simulat}],
gravitational instability [\ref{Korchagin-etal-2000!N1566}],
interactions with another galaxies [\ref{Tutukov-Fedorova-2006}],
hydrodynamical instabilities [\ref{Fridman-2008!UFN-instab}],
induced star formation in flocculent galaxies
[\ref{Berman-2003!Floc-simul}].

\begin{figure*}
 \vskip 0.\hsize \hskip 0.0\hsize
\centerline
{\includegraphics[width=0.5\hsize]%{empty.eps}%
{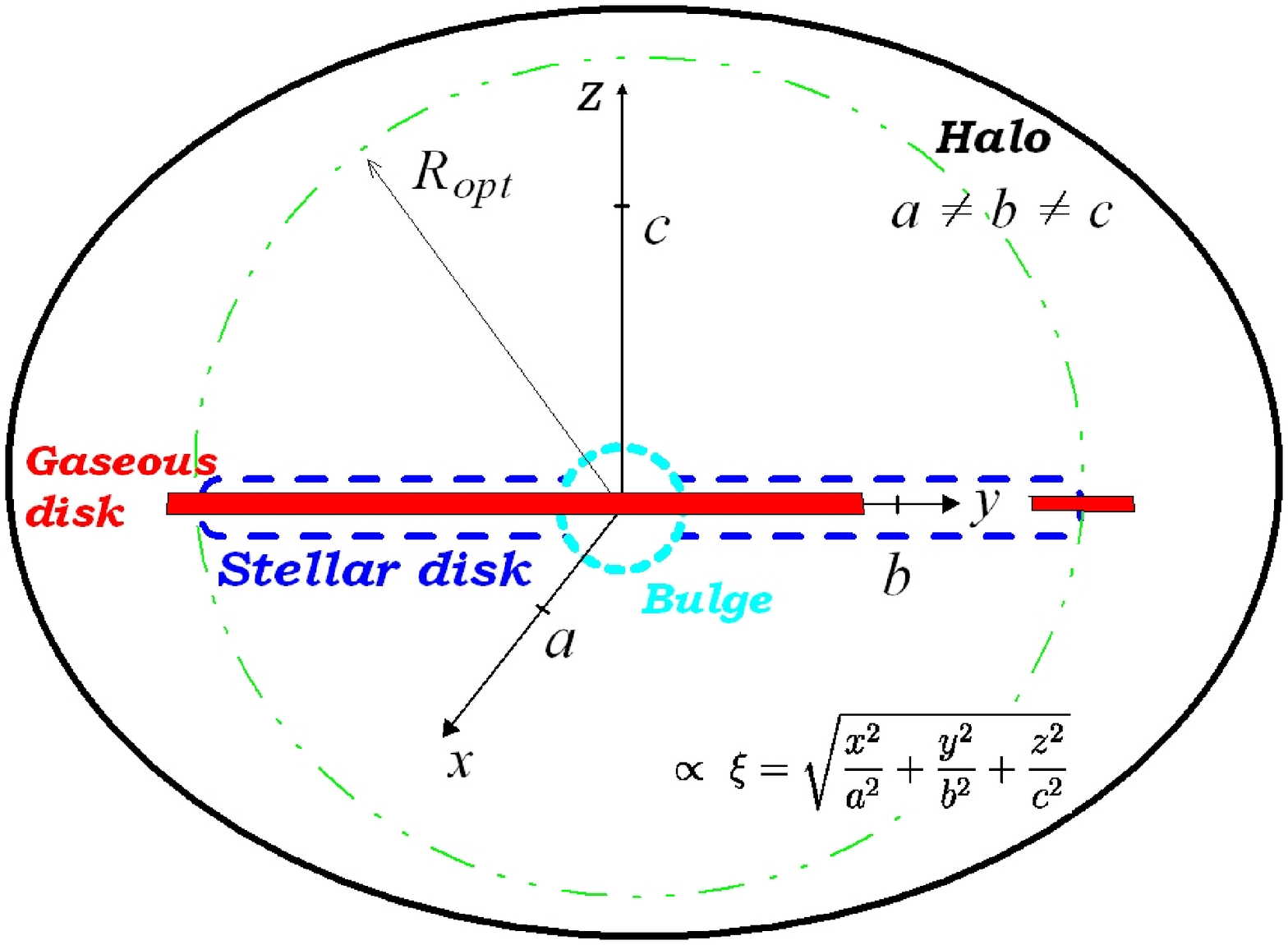}
}
\vskip 0.0\hsize \hskip 0.0\hsize \vbox{\hsize=0.999\hsize\caption
{ Galaxy disk submerged into triaxial halo}\label{Fig-haloshape}
}\vskip 0.0\hsize
\end{figure*}

  In this work we discuss the possibility of spiral structure formation in disks which are submerged into nonaxisymmetric dark halo (scales $a$, $b$, $c$ are different (fig.~\ref{Fig-haloshape}) ).

 In numerical CDM models dark halo structure significantly differs from central symmetry [\ref{Allgood-etal-2006}, \ref{Bailin-Steinmetz-2005},  \ref{Bullock-2002}, \ref{Jing-Suto-2005}, \ref{Kuhlen-etal-2007}].
 Attitude halo scales $q=b/a$, $s=c/a$ can reach 0.5--0.8.

 The various observational data indicate a possibility of the triaxial shape of halo.
Important markers are
\begin{description}
    \item[$\triangleleft$] Stellar halo objects (RR Lyrae, G,F-dwarfs, SDSS)
    [\ref{Keller-etal-2008}, \ref{Newberg-Yanny-2005}, \ref{Newberg-Yanny-2006},
     \ref{Xu-etal-2006}].
    \item[$\triangleleft$] Tidal streams in Milky Way and in other galaxies
    (Sagittarius tidal stream, Sagittarius dSph [\ref{Johnston-etal-2005},
    \ref{Helmi-2004}], Virgo Over Density  [\ref{Keller-etal-2008},
    \ref{Newberg-etal-2007}], The Sagittarius Dwarf Galaxy [\ref{Law-Majewski-2010}]).
    \item[$\triangleleft$] Polar Rings [\ref{Whitemore-etal-1987},
    \ref{Iodice-etal-2003}, \ref{Reshetnikov-Sotnikova-2000},
\ref{Combes-2006}, \ref{Brosch-Moiseev-2010}].

    \item[$\triangleleft$] Hyper Velocity Stars (HVS) [\ref{Brown-etal-2008},
    \ref{Gnedin-etal-2005}, \ref{Gualandris-Zwart-2007}].

    \item[$\triangleleft$] Galactic warps [\ref{Jeon-etal-2009!Galactic-Warps-Triaxial-Halos}].

    \item[$\triangleleft$] Some features of galaxies  structure can give the information
about the shape DH [\ref{Olling-Merrifield-2000},
\ref{Narayan-etal-2005}, \ref{Kalberla-etal-2007},
\ref{O'Brien-etal-2010}, \ref{Valluri-etal-2010},
\ref{Roskar-etal-2010}].

\end{description}

The most important thing for our work is nonaxisymmetric distribution of dark matter in the disk plane(fig.~\ref{Fig-haloshape}).
We will describe this feature by parameter
\begin{equation}\label{eq-epsilon-halo}
    \varepsilon = 1 - \frac{b}{a} \,,
\end{equation}
which is limited as follows $0 < \varepsilon \lee 0.15$.
We shall consider the effect of the shape of dark halo on the dynamics of the disk for $\varepsilon> 0$.

\begin{center}
  \textbf{2. Dynamical models of stellar and gaseous disks}
\end{center}

We have considered dynamics of stellar and gaseous disks in
nonaxisymmetric halo separately for stars and gas. The basis of
model of collisionless stellar disk is N-body dynamical system and
Poisson equation for gravitation:
\begin{equation}\label{eq-model-stellar}
     \frac{d^2 \vec{r}_i}{dt} = \frac{\| \Phi}{\| \vec{r}} +
\vec{F}_{halo}(\vec{r}_i,t) + \vec{F}_{bulge}(\vec{r}_i) \,, \ \ \
\ \ i=1,2,...,N
\end{equation}
\begin{equation}\label{eq-Poisson}
    \Delta\Phi = \frac{\|}{r\,\|r}\left(
    r\,\frac{\|\Phi}{\|r}\right)+ \frac{\|^2
    \Phi}{r^2\,\|\varphi^2}+ \frac{\|^2\Phi}{\|z^2} = 4\pi
    G\varrho \,.
\end{equation}

 In case of gas disk we solve the classical hydrodynamics equations:
\begin{equation}\label{eq-hydrodynamics-rho}
     \frac{\| \varrho}{\| t} + \vec{\nabla}(\varrho \vec{\bf u}) =
     0\,,
\end{equation}
\begin{equation}\label{eq-hydrodynamics-u}
 \frac{d(\varrho\vec{\bf u})}{dt} = - \vec{\nabla}{P} - \varrho
\,\vec{\nabla}\Phi_{sum} \,,
\end{equation}
\begin{equation}\label{eq-hydrodynamics-E}
 \varrho \frac{d}{dt}\left( \frac{E}{\varrho} \right) -
\frac{P}{\varrho}\frac{d\varrho}{dt} = 0 \,.
\end{equation}

 We used 2 numerical methods in our simulations:
 \begin{itemize}
    \item TVD MUSCL method
    \item Lagrangian method SPH with $h\ne \textrm{const}$ ($h$ is kernel scale) [\ref{Monaghan-sph}].
 \end{itemize}

We also used two rigid models of the nonaxisymmetric halo.
First is analog of potential of pseudo isothermal halo:
\begin{equation}\label{eq-pseudo-isothermal-halo}
        \Psi_h(x,y,z) = 4\pi G \varrho_{h0} a^2\cdot \Big\{
\ln(\xi) +
\end{equation}
$$
 \left. + \frac{{\rm arctg(\xi)}}{\xi} + \frac{1}{2} \ln
\frac{1+\xi^2}{\xi^2}
    \right\} \,,
$$
where $\Oo\xi=\sqrt{ \frac{x^2}{a_x^2} + \frac{y^2}{a_y^2} +
\frac{z^2}{a_z^2}}$.
 In the case of equal values of the scales $a = b = c$, we have a usual
model of symmetric pseudo isothermal halo.
 And the second halo model is analog of model by Navarro, Frenk \& White
 [\ref{Navarro-etal-1997}]:
\begin{equation}\label{eq-analog-NWF-halo}
    \varrho_h(x,y,z) = \frac{\varrho_{h0}}{\xi\,(1+\xi)} \,.
\end{equation}
That kinds of potentials reproduce flat rotation curve with many values of their parameters.

 We took into account solid-body rotation of these potentials with angular velocity $\Omega_{halo}$.
 Halo rotation is slower than rotation of disk  on periphery~($\Omega_{disk}$):
 \begin{equation}\label{eq-rotation-halo}
    \Omega_{halo} < \Omega_{disk}(R_{opt}) \,.
\end{equation}

\begin{center}
  \textbf{3. Stellar disk dynamics}
\end{center}

\begin{figure*}
\centerline
{\includegraphics[width=0.5\hsize]%{empty.eps}%
{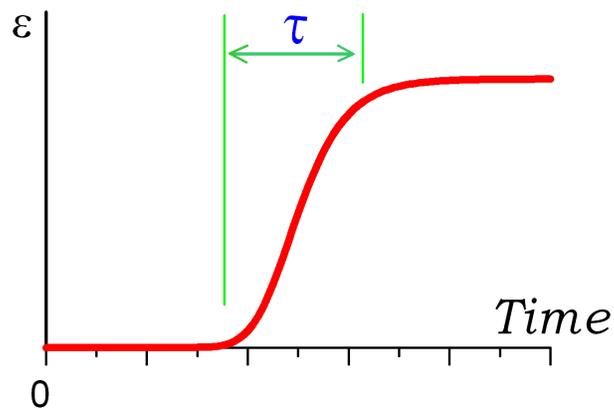}
}
\vskip  0.0\hsize \hskip 0.0\hsize
\vbox{\hsize=0.999\hsize\caption {  Nonaxissymmetric index as
function of time}\label{Fig-switch-halo}}
\end{figure*}

Our simulations start with axisymmetrical halo $\varepsilon(t=0) = 0$. Then halo became
nonaxisymmetric during some time interval $\tau$, which are 1--3 periods of disk rotation (fig.~\ref{Fig-switch-halo}).

\begin{figure*}
 \vskip 0.\hsize \hskip -0.0\hsize
\centerline {\includegraphics[width=0.5\hsize]{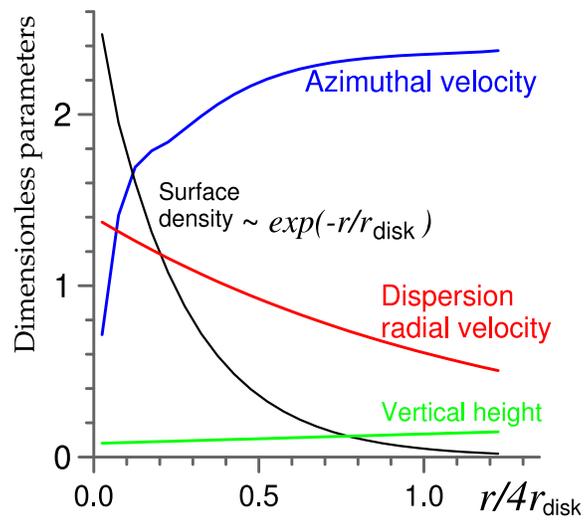} }
\vskip  0.0\hsize \hskip 0.0\hsize
  \vbox{\hsize=0.999\hsize
  \caption { Initial distributions of the basic parameters of the stellar disk}\label{Fig-initstar} }\vskip 0.0\hsize
\end{figure*}

\begin{figure*}
 \vskip -0.\hsize \hskip -0.0\hsize
\centerline {\includegraphics[width=0.5\hsize]{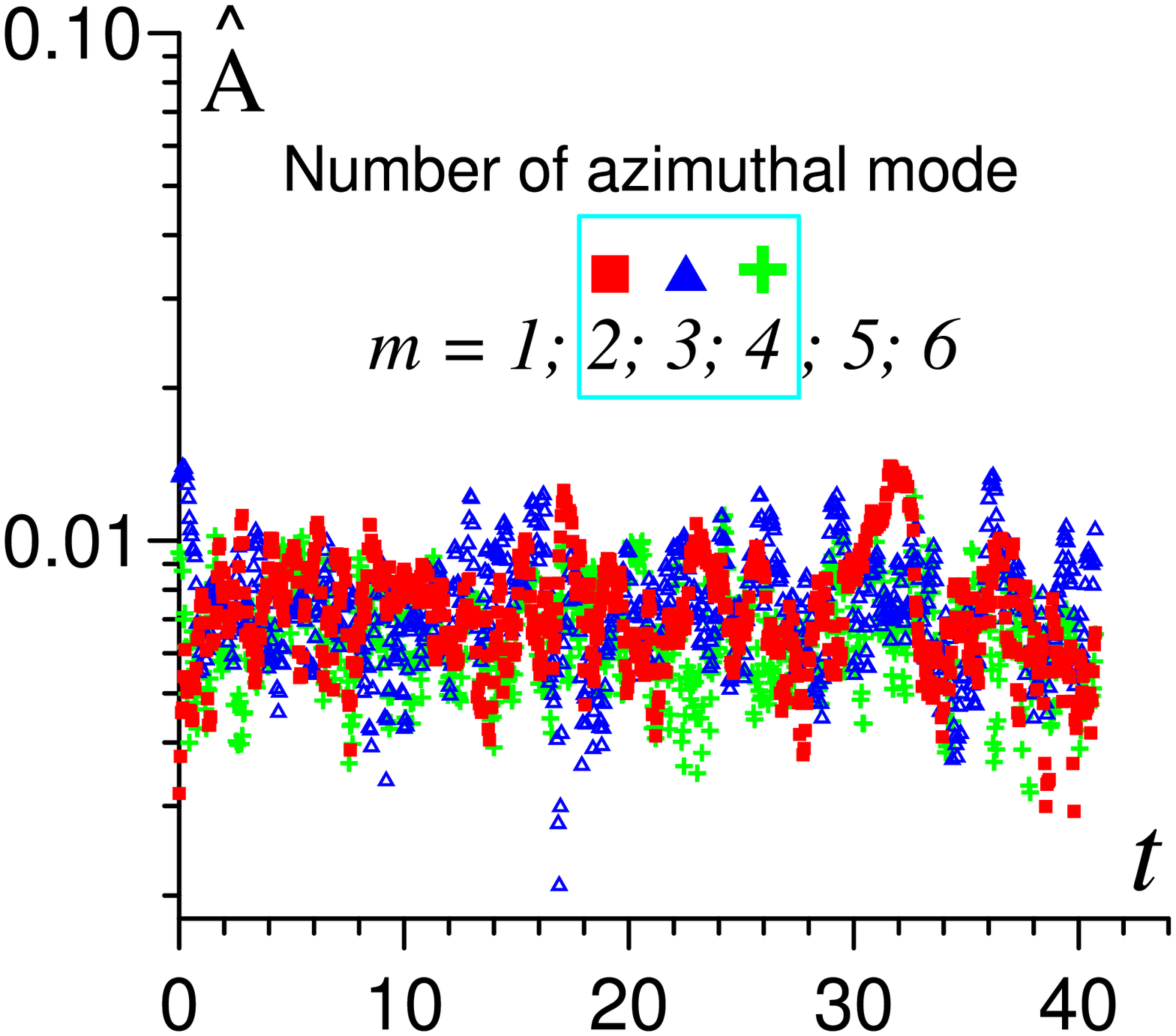} \ \ \
} \centerline {\includegraphics[width=0.5\hsize]{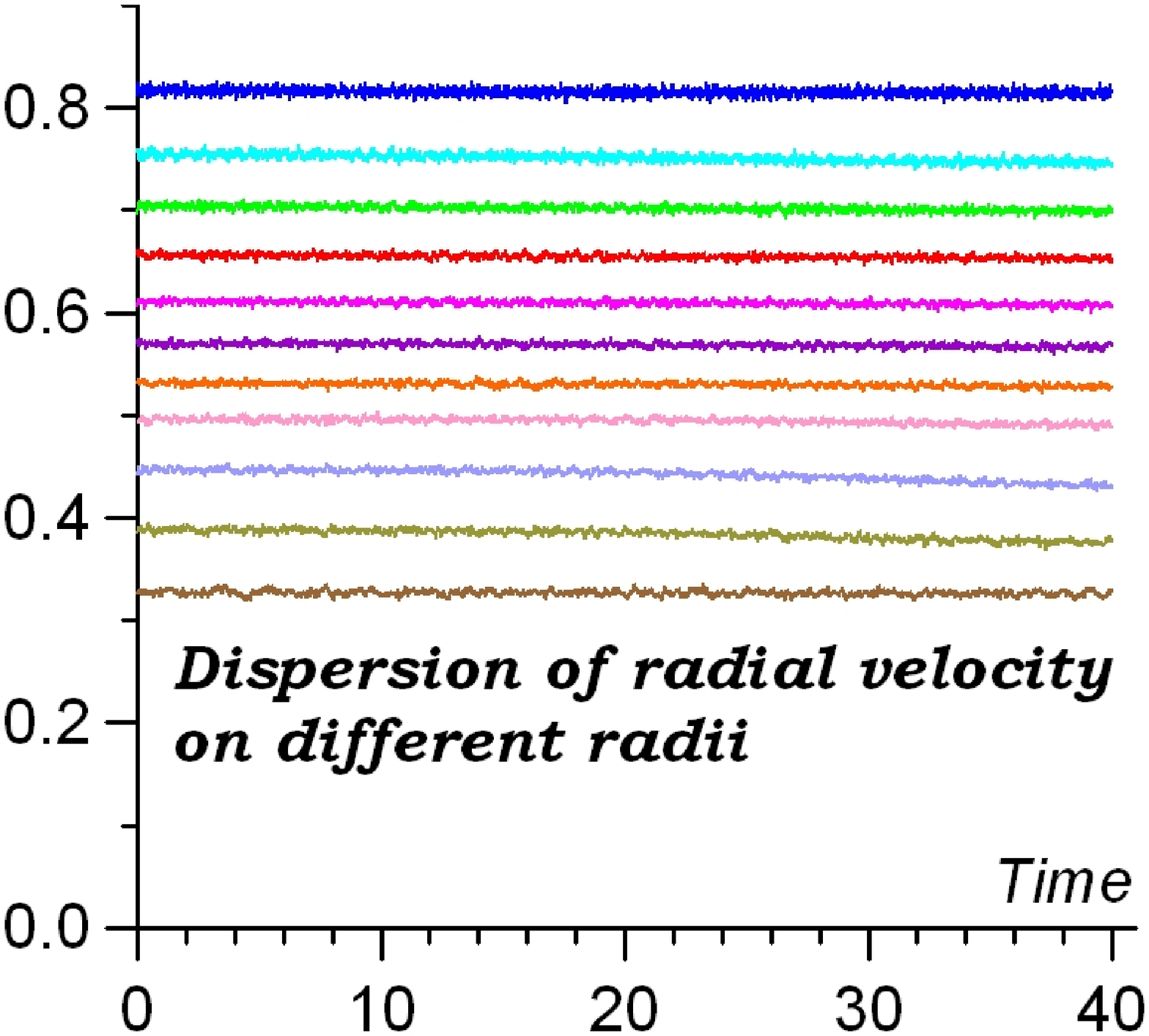}}

  \vbox{\hsize=0.999\hsize
  \caption { Evolution of stellar disk parameters in model with the axisymmetrical halo are shown. There is no growth of perturbations.}\label{Fig-sumfur0} }\vskip 0.0\hsize
\end{figure*}

In the fig.~\ref{Fig-initstar} radial profiles of
surface density $\sigma$, rotation curve  $V$, dispersion of
radial velocities of particles $c_r$, disk vertical scale are
shown for the stellar disk. These distributions ensure the
gravitational stability of the disk.
 Our initial radial distribution is stable because we use Toomre parameter $Q_T=c_r/(3.36 G\sigma/\varkappa)$ exceeding 2.
 All parameters do not change with time in case of axisymmetrical halo (fig.~\ref{Fig-sumfur0}).

\begin{figure*}
\centerline
{\includegraphics[width=0.9\hsize]%{empty.eps}%
{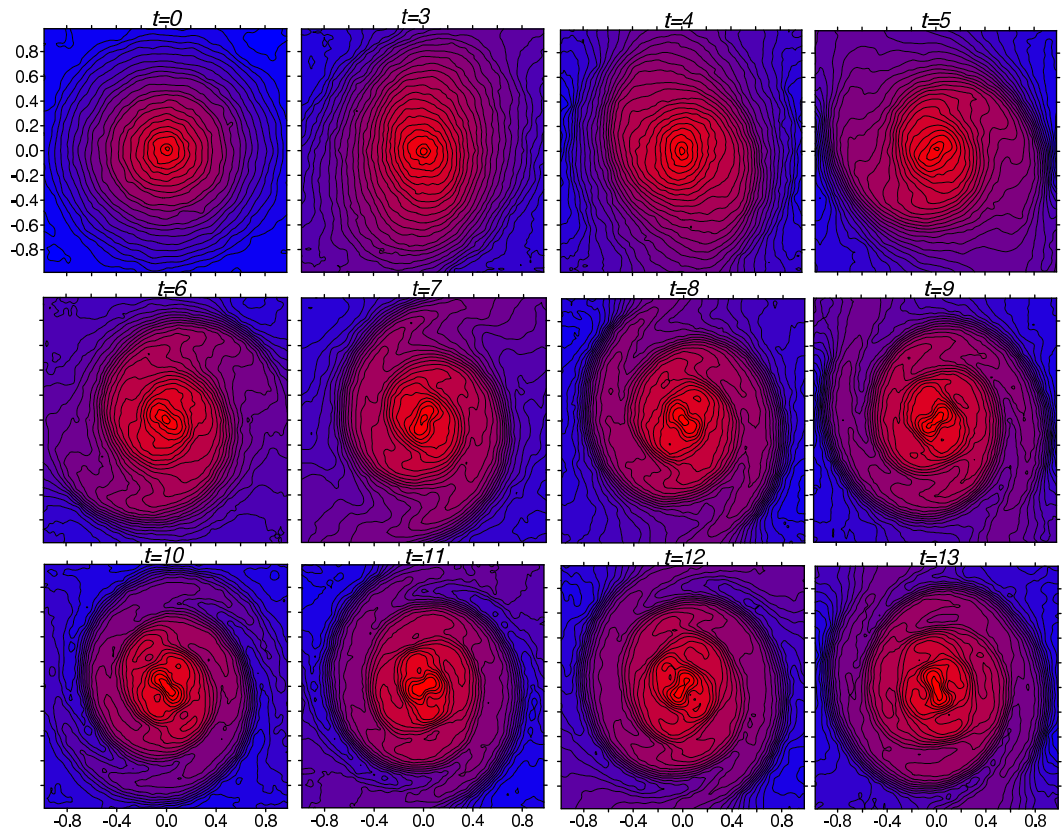}
}
 \vskip  -0.1\hsize \hskip 0.0\hsize
  \vbox{ \hsize=0.999\hsize \caption { The logarithm of surface density in stellar disk model }\label{Fig-stellar-disk} }
\end{figure*}

 The situation changes in the case of nonaxisymmetric halo. Fig.~\ref{Fig-stellar-disk} show the distributions of the
logarithm of the surface density in the disk at different times.

\begin{figure*}
\vskip 0.\hsize \hskip 0.0\hsize
 \centerline
{\includegraphics[width=0.5\hsize]%{empty.eps}%
{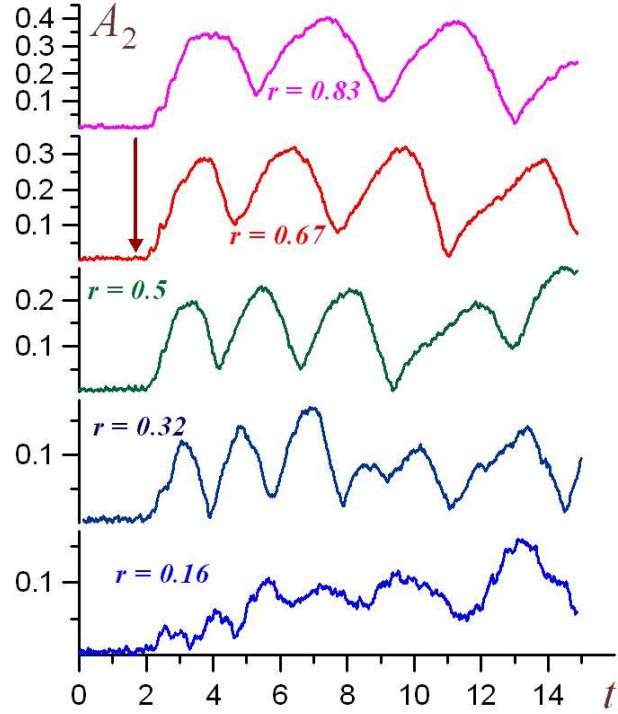}
}
\vskip  0.0\hsize \hskip 0.0\hsize \vbox{\hsize=0.999\hsize
\caption { Evolution of Fourier harmonics amplitude for two-arms
mode in different radii Value $r=1$ corresponds
$4r_d$.}\label{Fig-AmpFour-stars} }\vskip 0.0\hsize
\end{figure*}

\begin{figure*}
 \vskip 0.\hsize \hskip -0.0\hsize
\centerline
{\includegraphics[width=0.5\hsize]%{empty.eps}%
{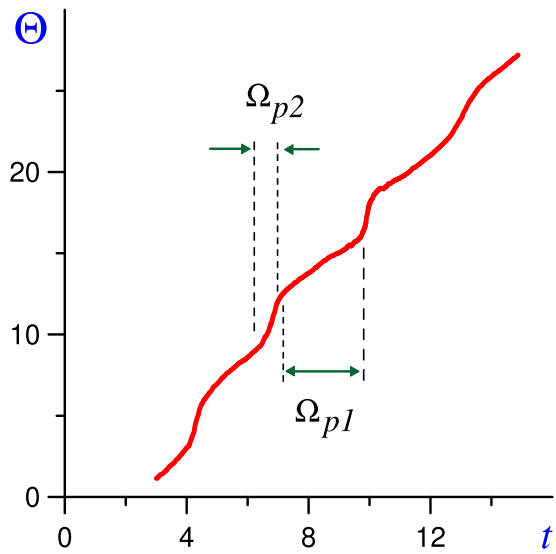}
}
\vskip  0.0\hsize \hskip 0.0\hsize
  \vbox{\hsize=0.999\hsize
  \caption { Evolution of Fourier harmonics phase $\Theta$ for two-arms mode on fixed radius}\label{Fig-Phase-stars} }\vskip 0.0\hsize
\end{figure*}

The amplitudes of Fourier-harmonics $ m = 2 $ at various radii are
more complicated.
 We observe quasiperiodic changes of Fourue-amplitudes for two-arm
 wave (fig.~\ref{Fig-AmpFour-stars}). As the result a rotation velocity of spiral pattern is
  non-stationary (fig.~\ref{Fig-Phase-stars}).

  Angular velocity of rotation $\Omega_p$ is determined by spiral wave phase $\Theta$:

\begin{equation}\label{eq-Omega-Phase}
    \Omega_p = \frac{d\Theta}{dt} \,.
\end{equation}

Two stages of disk dynamics recognized well. Slow rotation $\Omega_{p1}$ related with dynamic of central region and fast rotation $\Omega_{p2}$ corresponded with outer disk part.

\begin{center}
  \textbf{4. Gaseous disk dynamics} %Äèíàìèêà ãàçîâîãî äèñêà}
\end{center}

\begin{figure*}
 \vskip 0.\hsize \hskip 0.0\hsize
            \includegraphics[width=0.25\hsize]{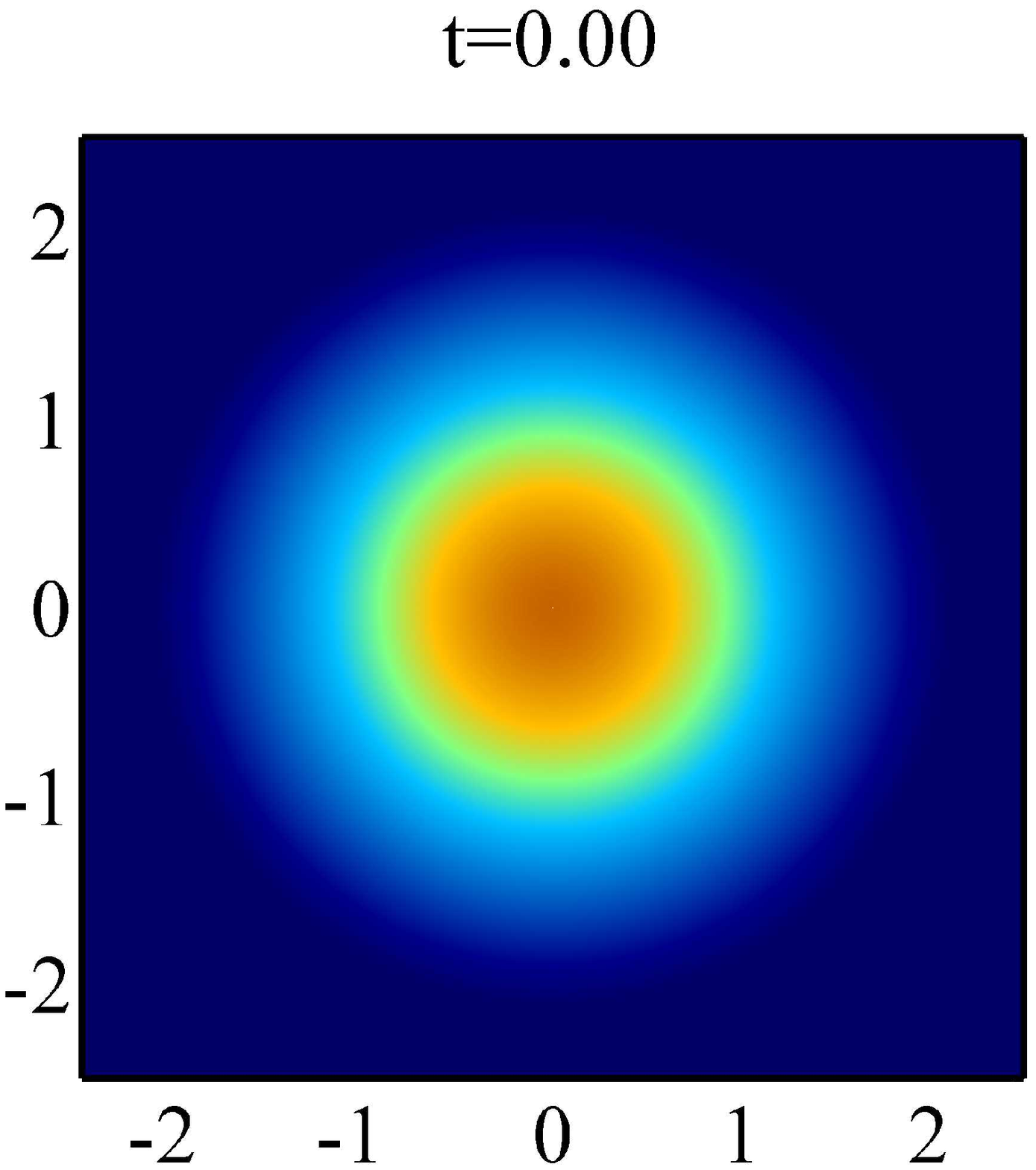}
 \vskip -0.232\hsize \hskip 0.250\hsize
            \includegraphics[width=0.25\hsize]{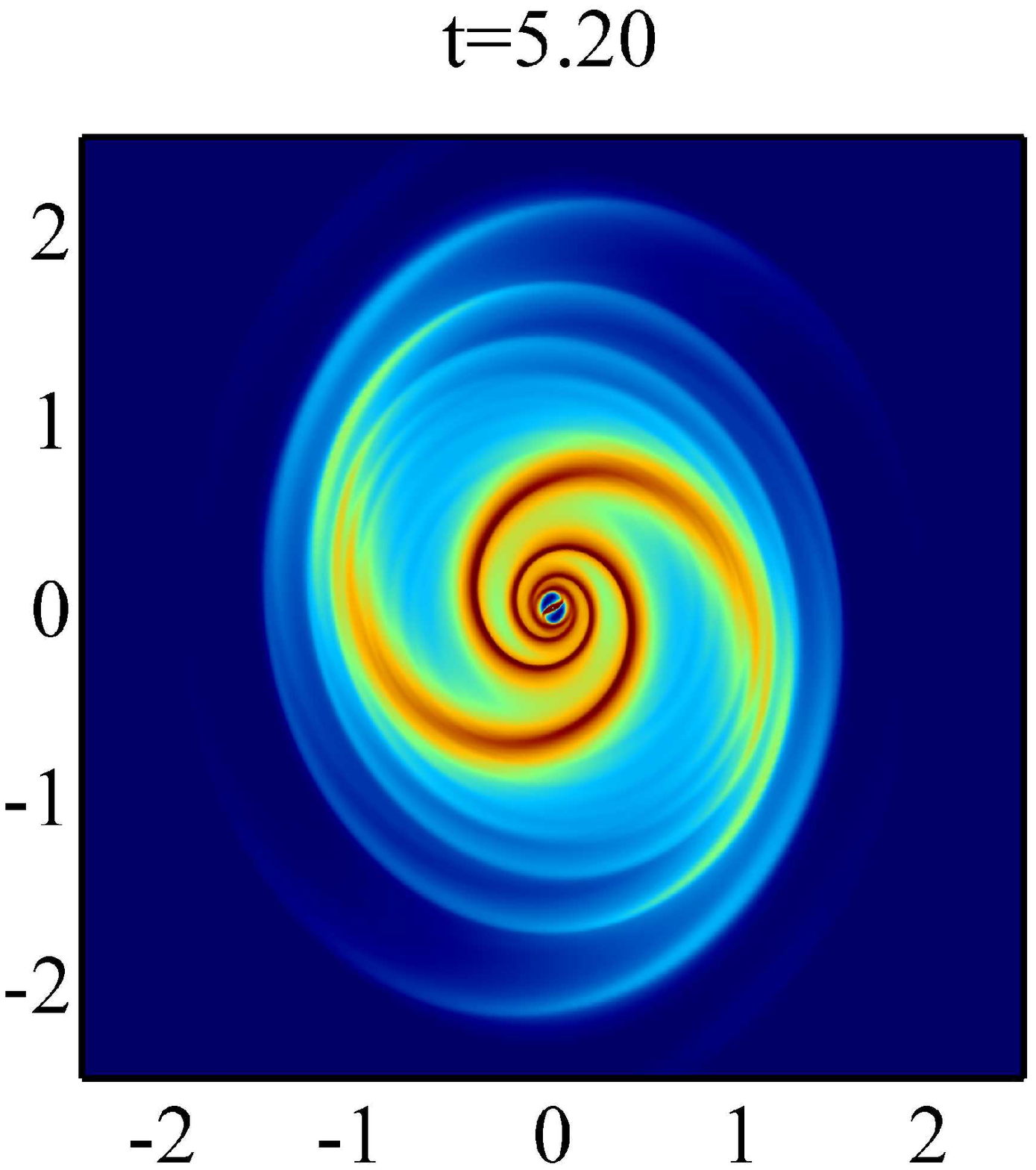}
 \vskip -0.232\hsize \hskip 0.50\hsize
            \includegraphics[width=0.25\hsize]{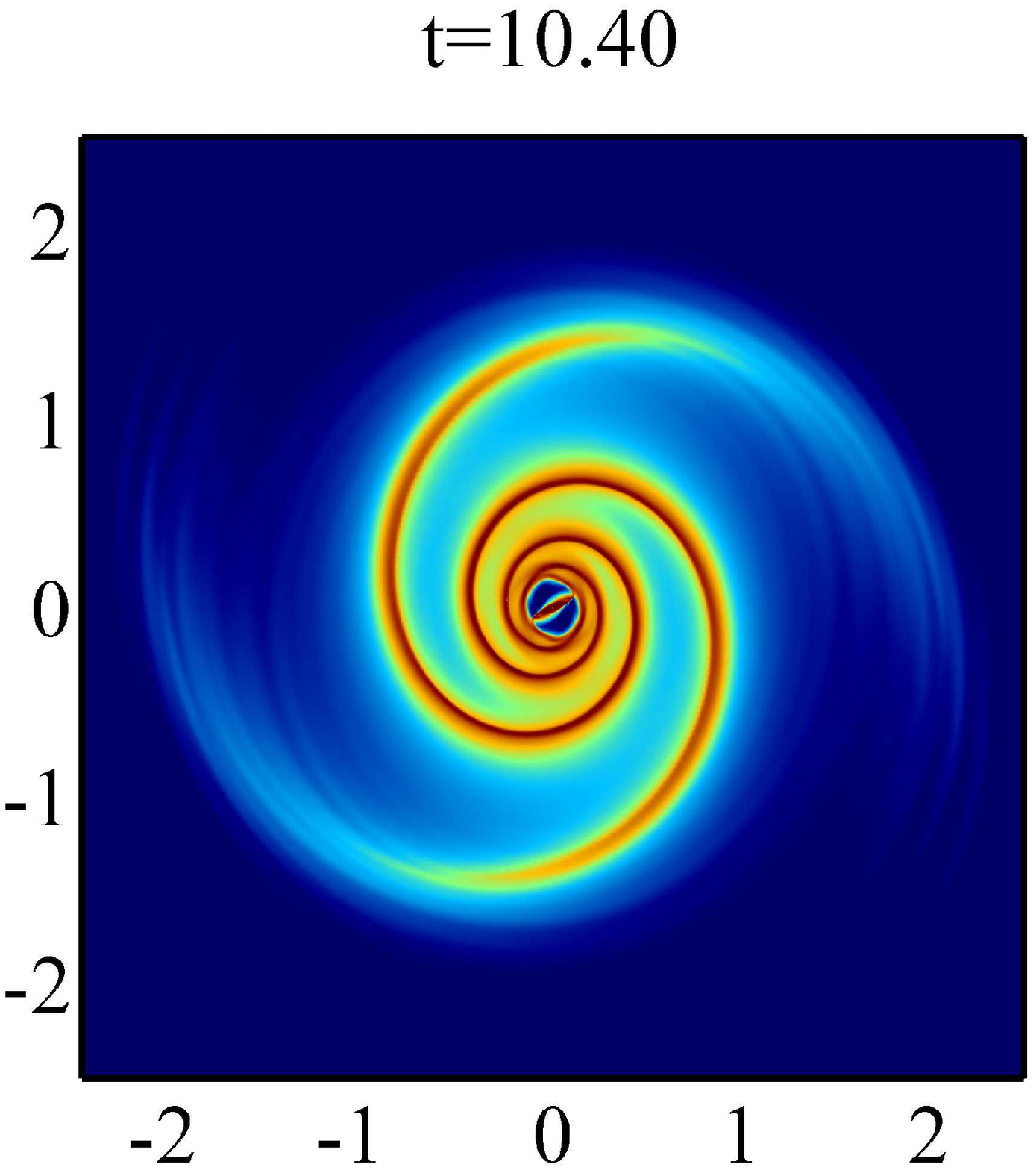}
 \vskip -0.232\hsize \hskip 0.750\hsize
            \includegraphics[width=0.25\hsize]{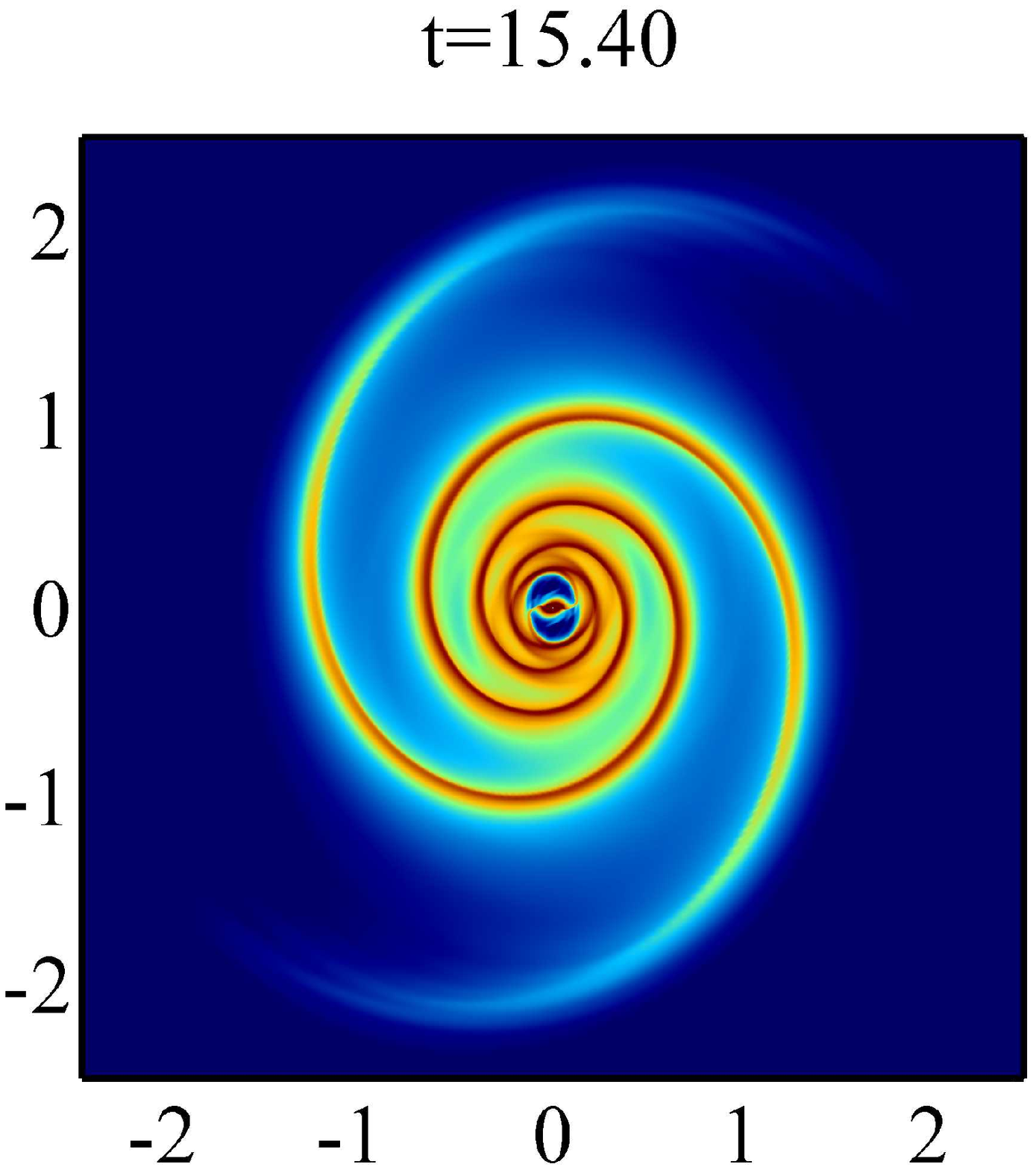}
 \vskip 0.\hsize \hskip 0.0\hsize
            \includegraphics[width=0.25\hsize]{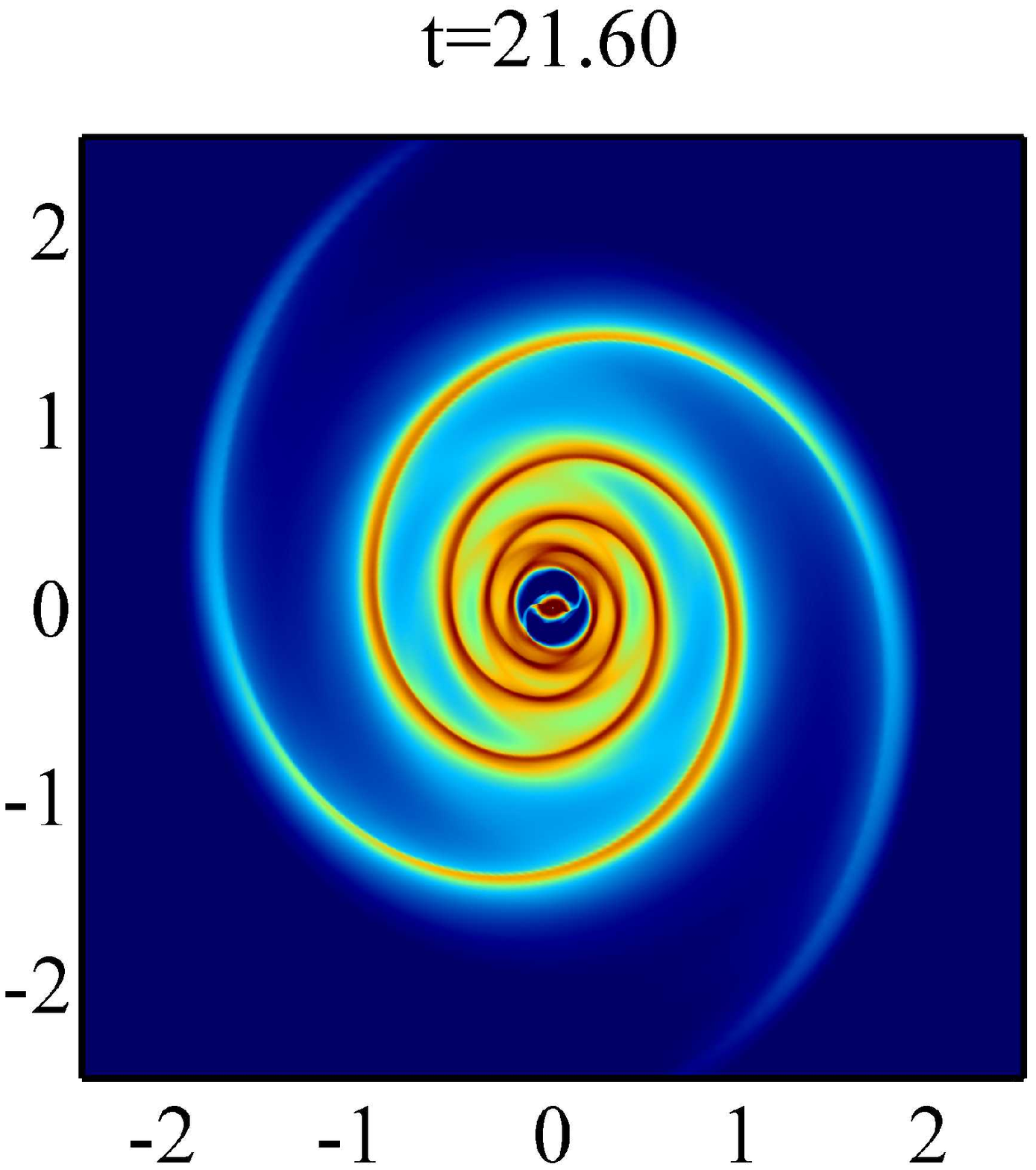}
 \vskip -0.232\hsize \hskip 0.250\hsize
            \includegraphics[width=0.25\hsize]{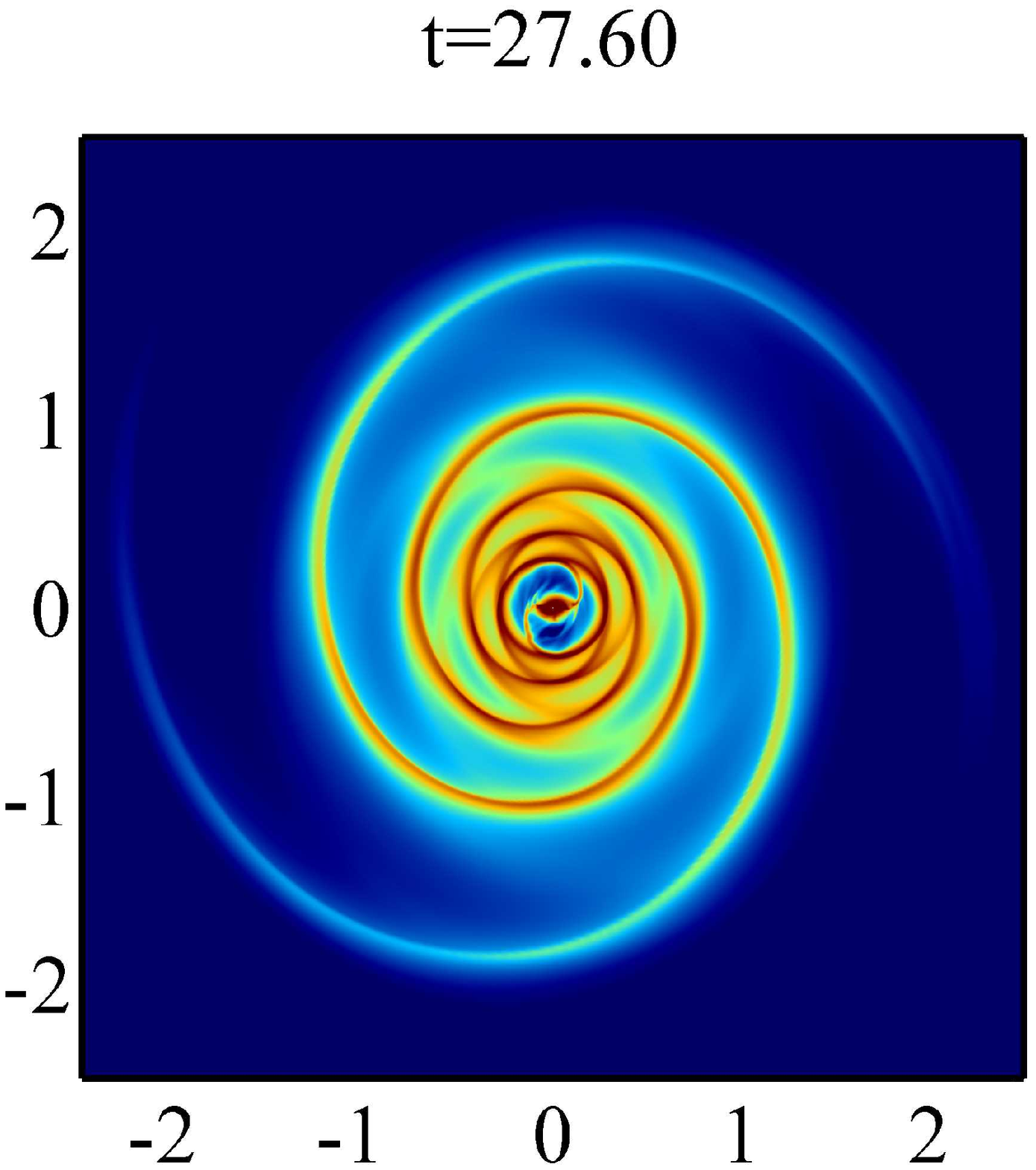}
 \vskip -0.232\hsize \hskip 0.50\hsize
            \includegraphics[width=0.25\hsize]{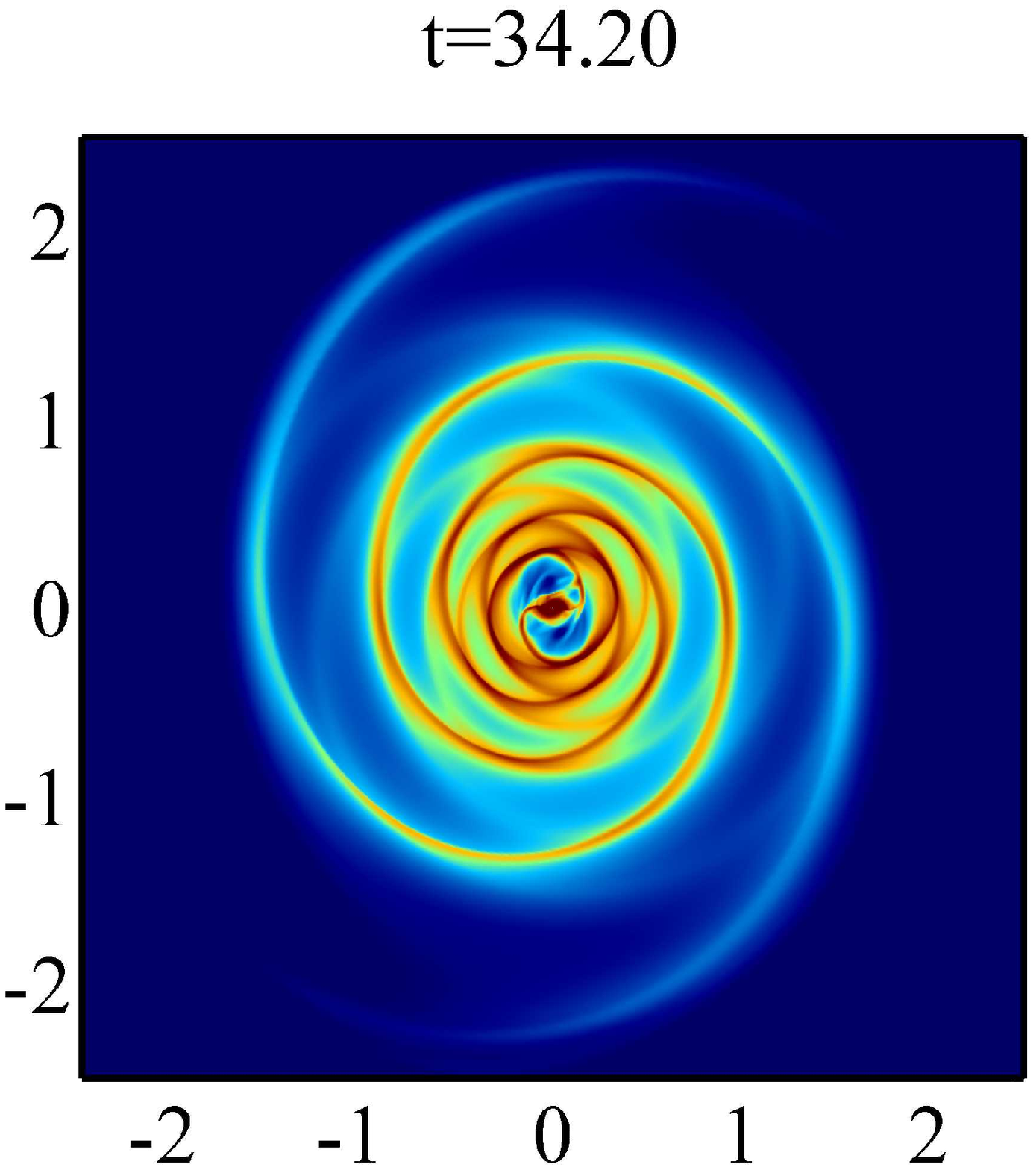}
 \vskip -0.232\hsize \hskip 0.750\hsize
            \includegraphics[width=0.25\hsize]{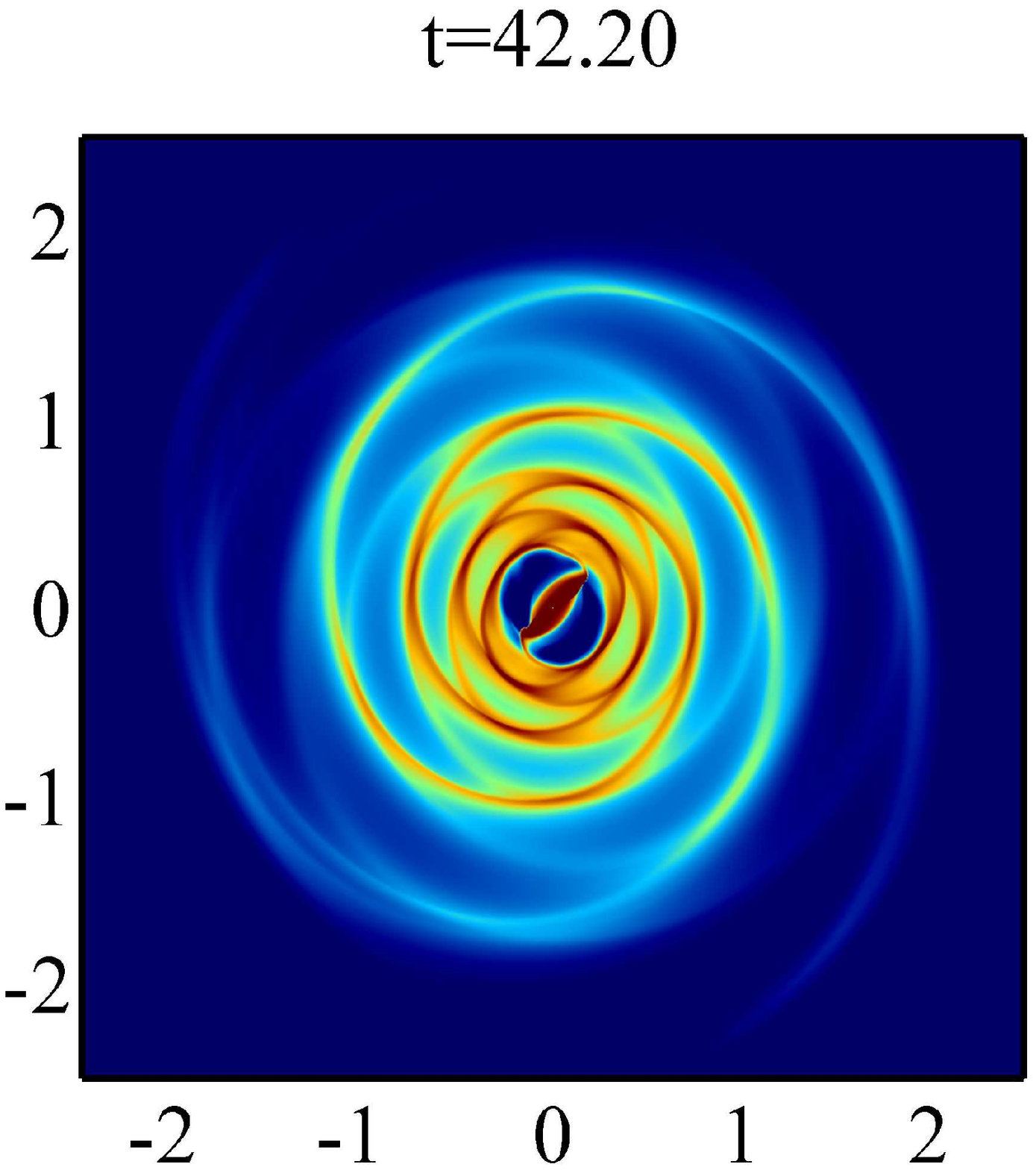}
\vskip  0.0\hsize \hskip 0.0\hsize
  \vbox{\hsize=0.999\hsize
  \caption { Evolution of gaseous disk submerged into nonaxissymmetric halo }\label{Fig-evol-gas} }\vskip 0.0\hsize
\end{figure*}

\begin{figure*}
 \vskip 0.\hsize \hskip 0.0\hsize
 \centerline
{
 \includegraphics[width=0.5\hsize]{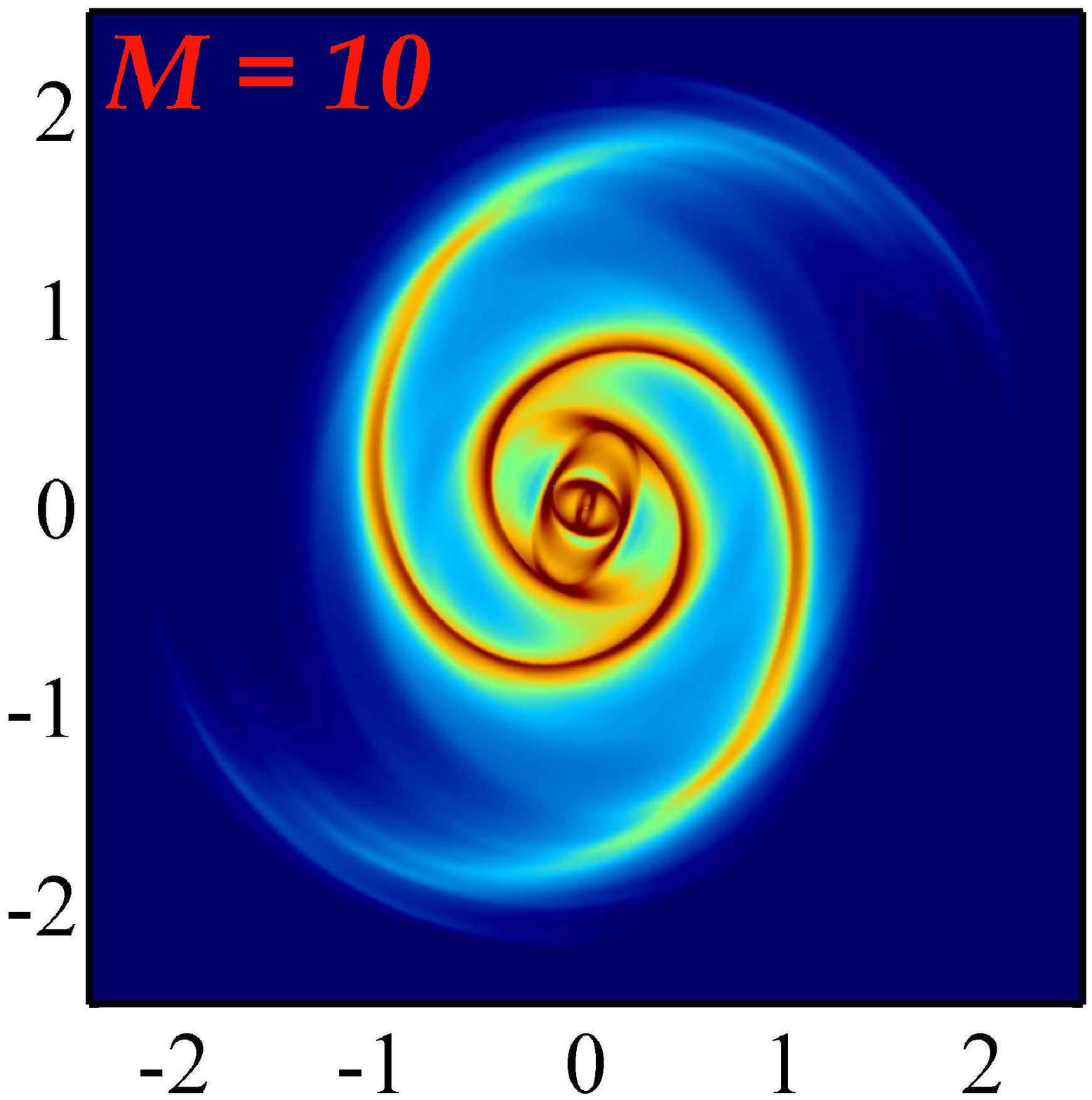}
 \includegraphics[width=0.5\hsize]{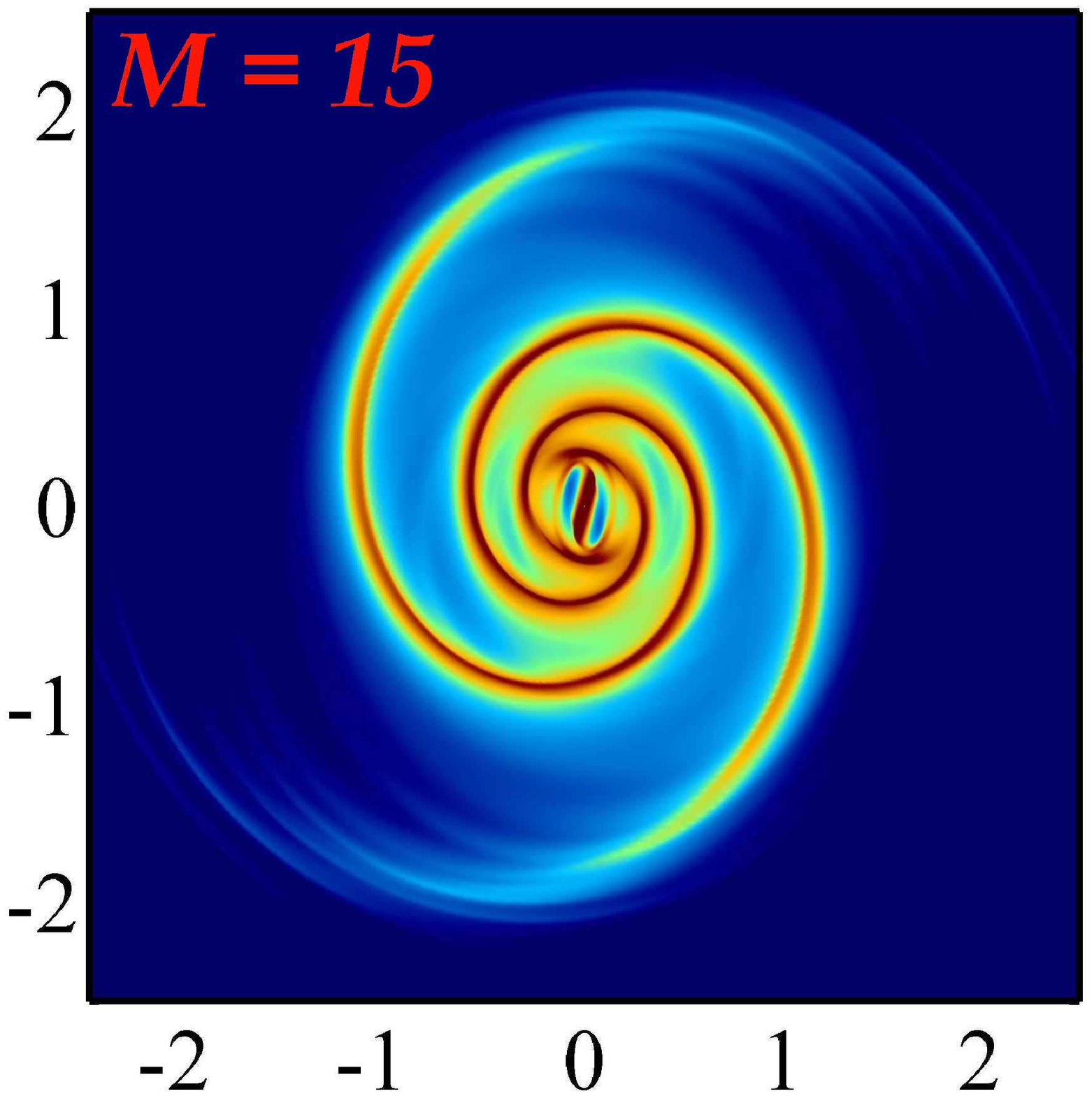}
 }
 \centerline
{
 \includegraphics[width=0.5\hsize]{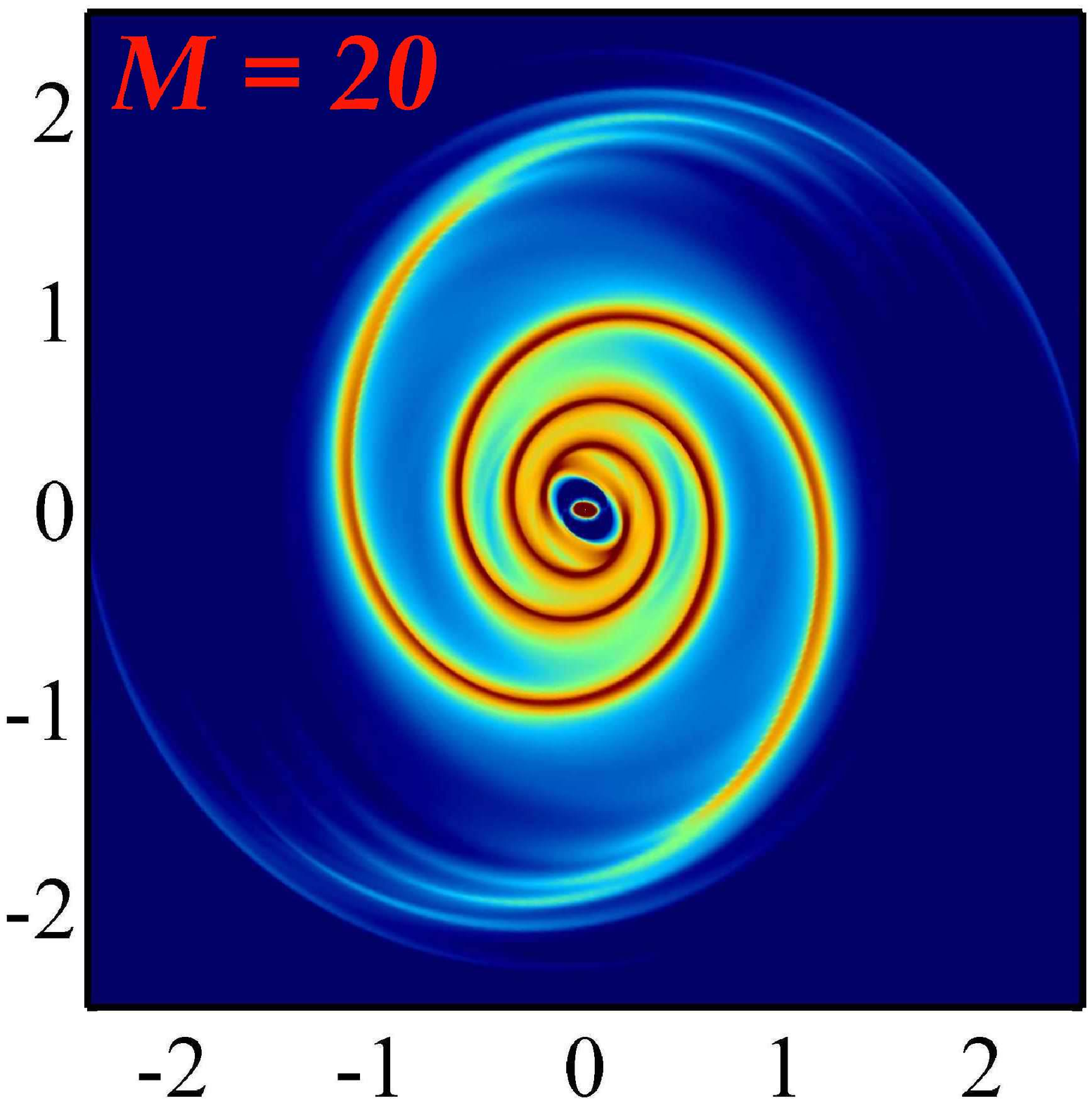}
 \includegraphics[width=0.5\hsize]{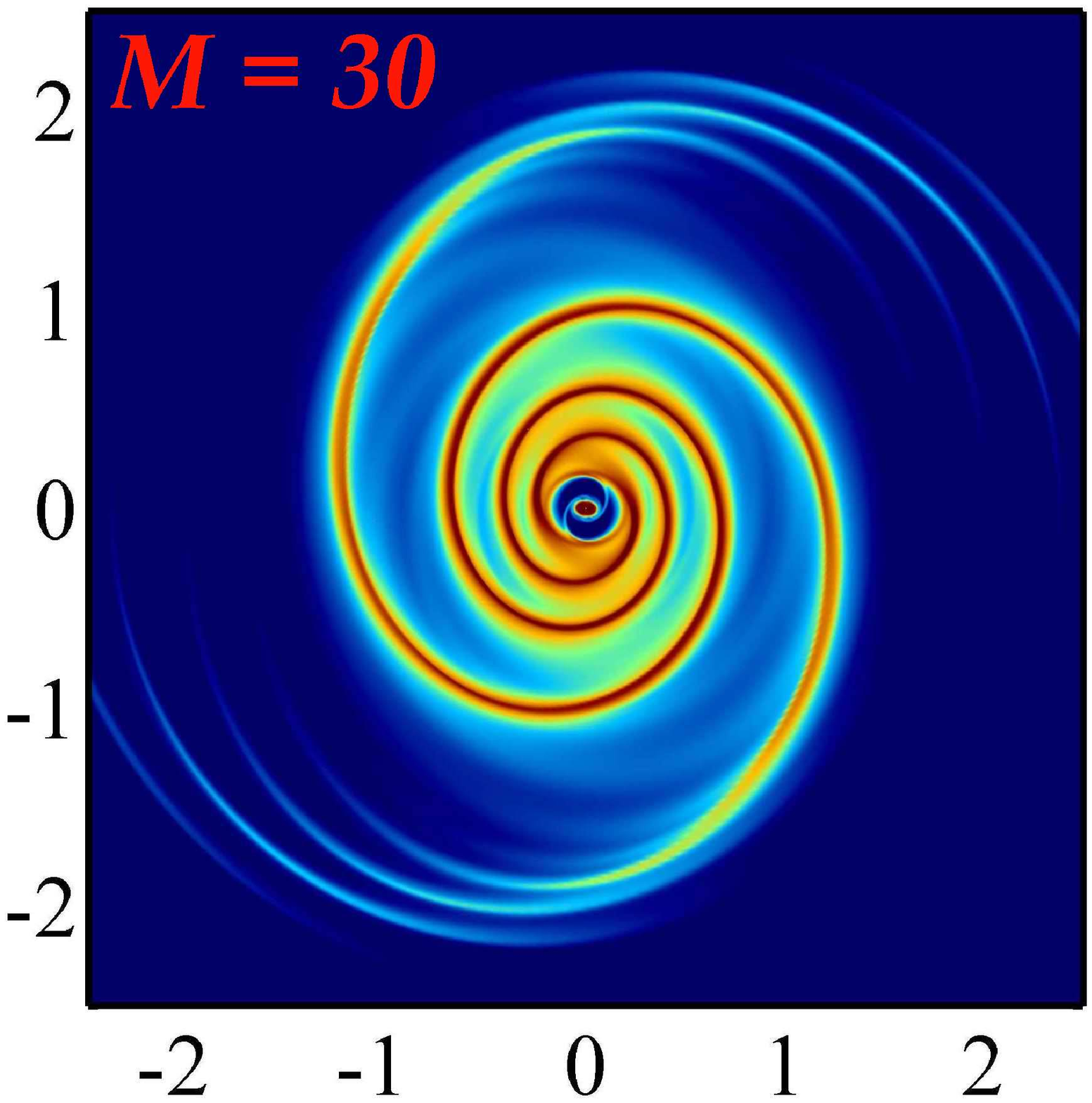}
 }

\vskip  0.0\hsize \hskip 0.0\hsize
  \vbox{\hsize=0.999\hsize
  \caption {Distributions of surface density in gaseous disks for different values of Mach number $M$ }\label{Fig-mach} }\vskip 0.0\hsize
\end{figure*}

 In fig.~\ref{Fig-evol-gas} the typical evolution of a gas disk is shown.
 It is the logarithm of the surface density in different times.
 The formation of two-arm trailing  spiral structure is observed.
 For typical conditions perturbations grow to nonlinear stage during 1-3 circulation and form system of shocks. Analysis of gas flow structure demonstrates it's difficult character. We also observe different shock waves and regions with share flow. Morphology of spiral structures depends on rotation curve $V(r)$, distribution of density in dark halo $\varrho_h(r,\varphi)$, radial distribution of stellar disk density $\sigma(r)$ and the sound speed. Specifically in central part of the disk often liding spirals forms. On periphery this spirals always become lagging spirals. Also liding spirals can produce $\Theta$-like structures (some times nested $\Theta$-like structures, for example see fig.\ref{Fig-mach} with $M=10$).

\begin{figure*}
 \vskip 0.\hsize \hskip 0.0\hsize
\centerline
{\includegraphics[width=0.4\hsize]%{empty.eps}%
{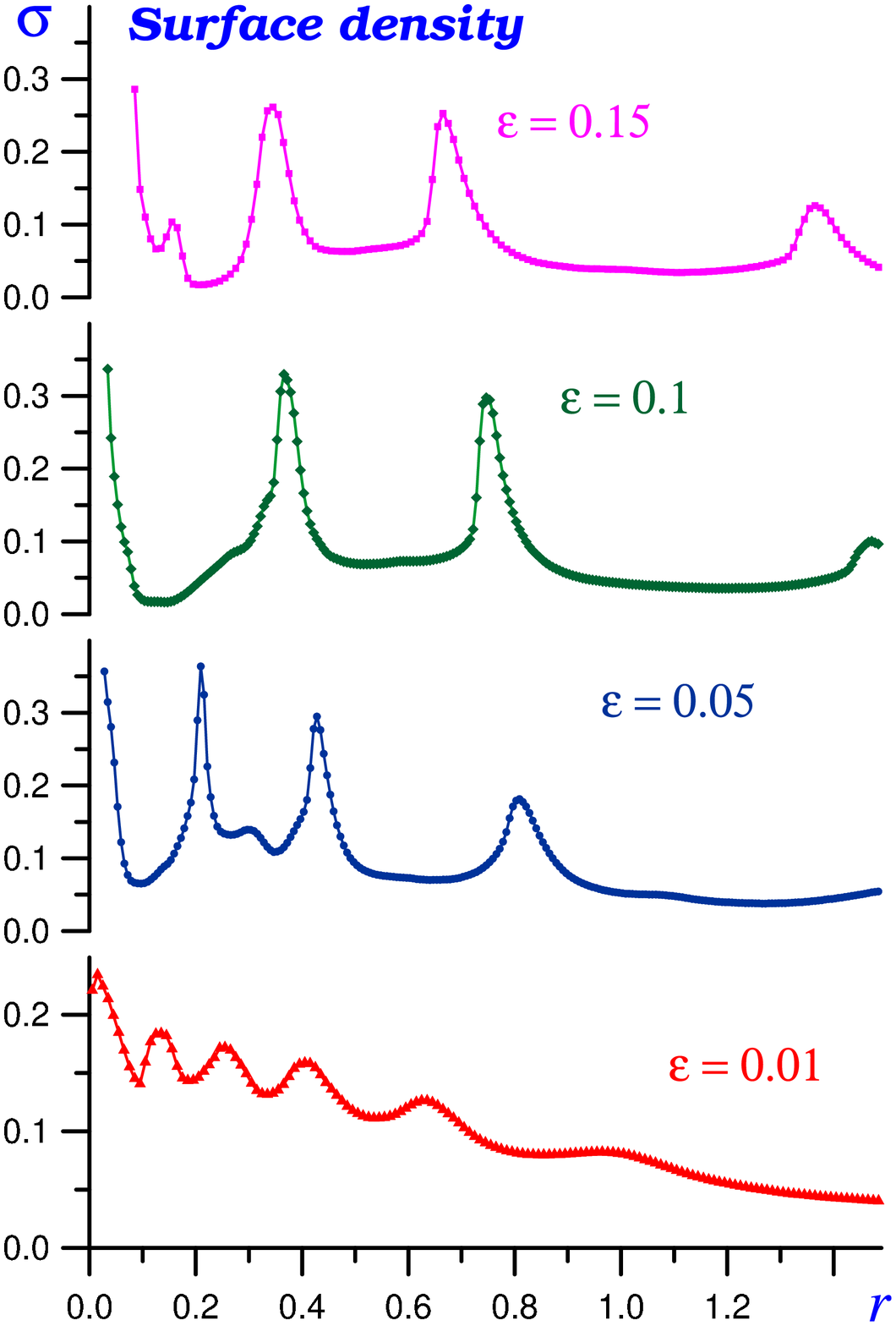} }
 \centerline
{
 \includegraphics[width=0.3\hsize]{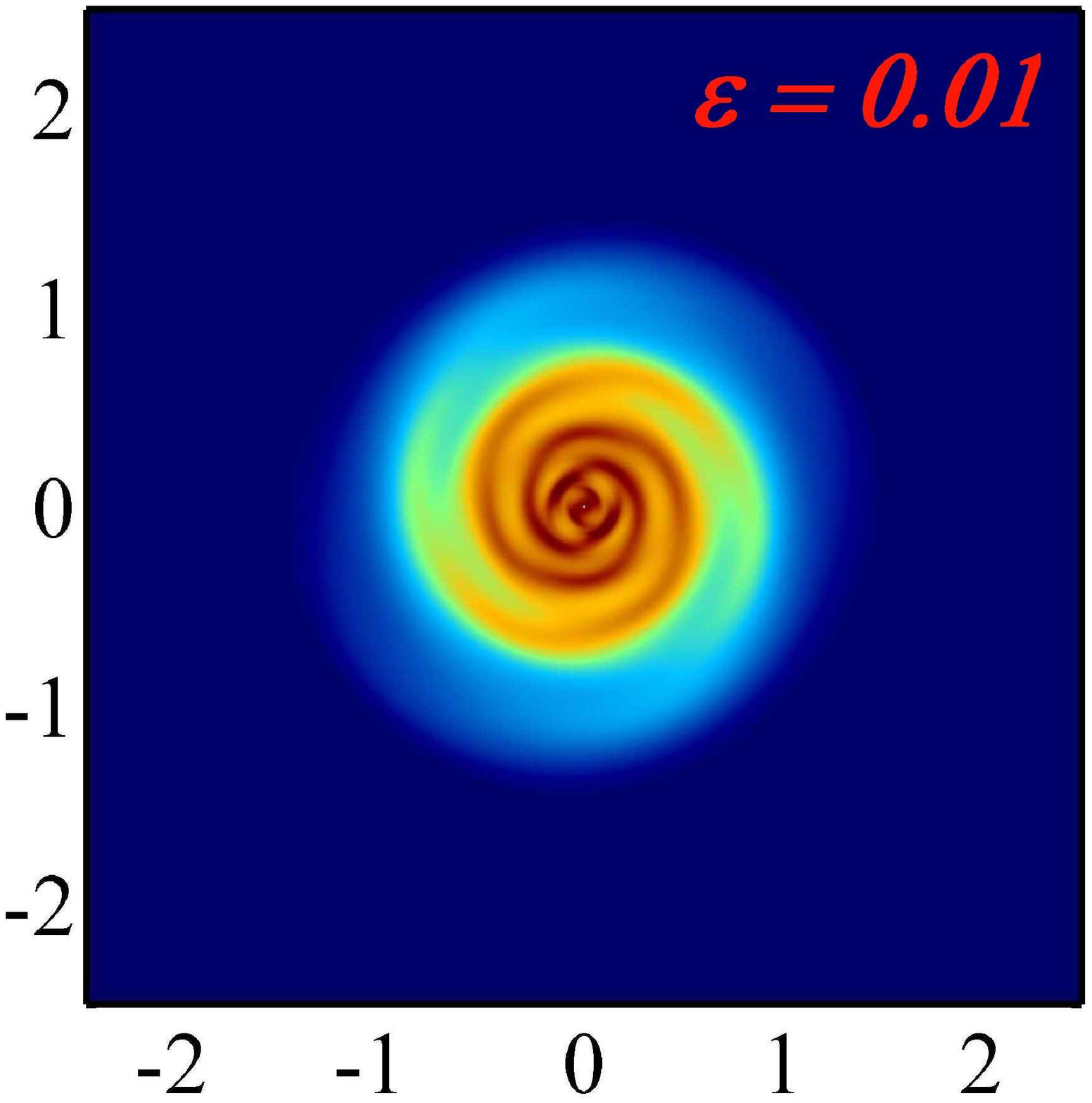}
 \includegraphics[width=0.3\hsize]{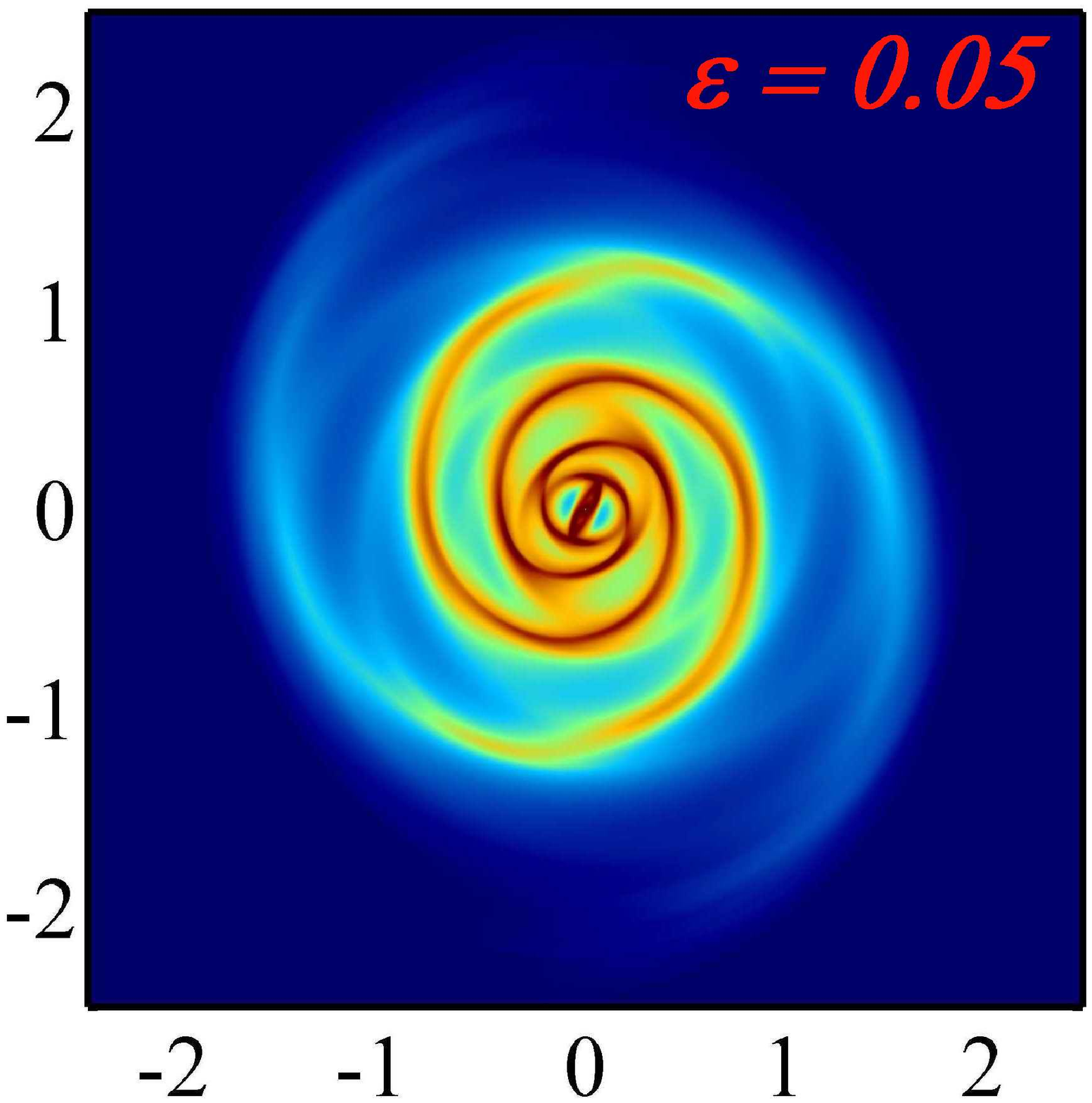}
 }
 \centerline
{
 \includegraphics[width=0.3\hsize]{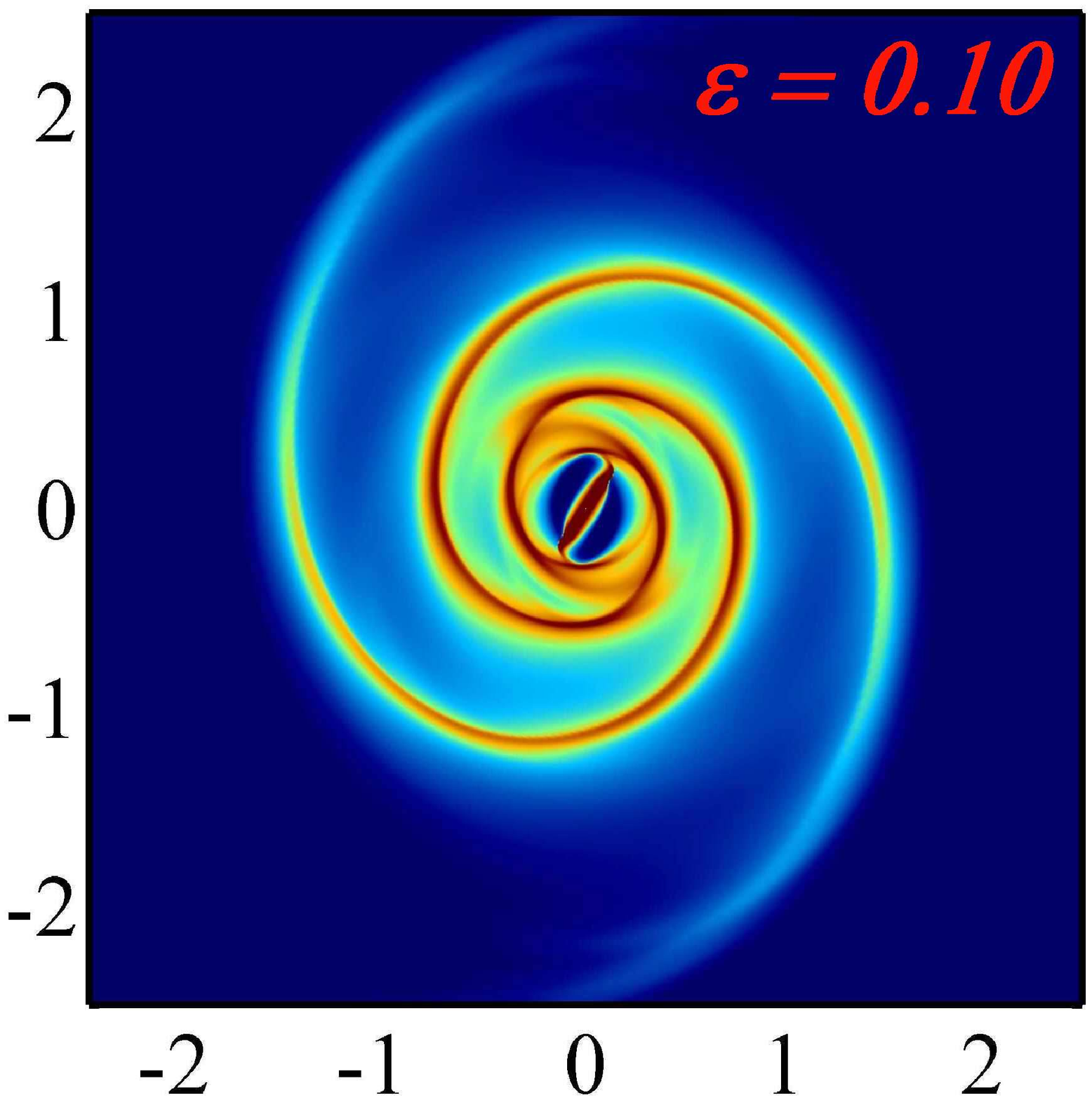}
 \includegraphics[width=0.3\hsize]{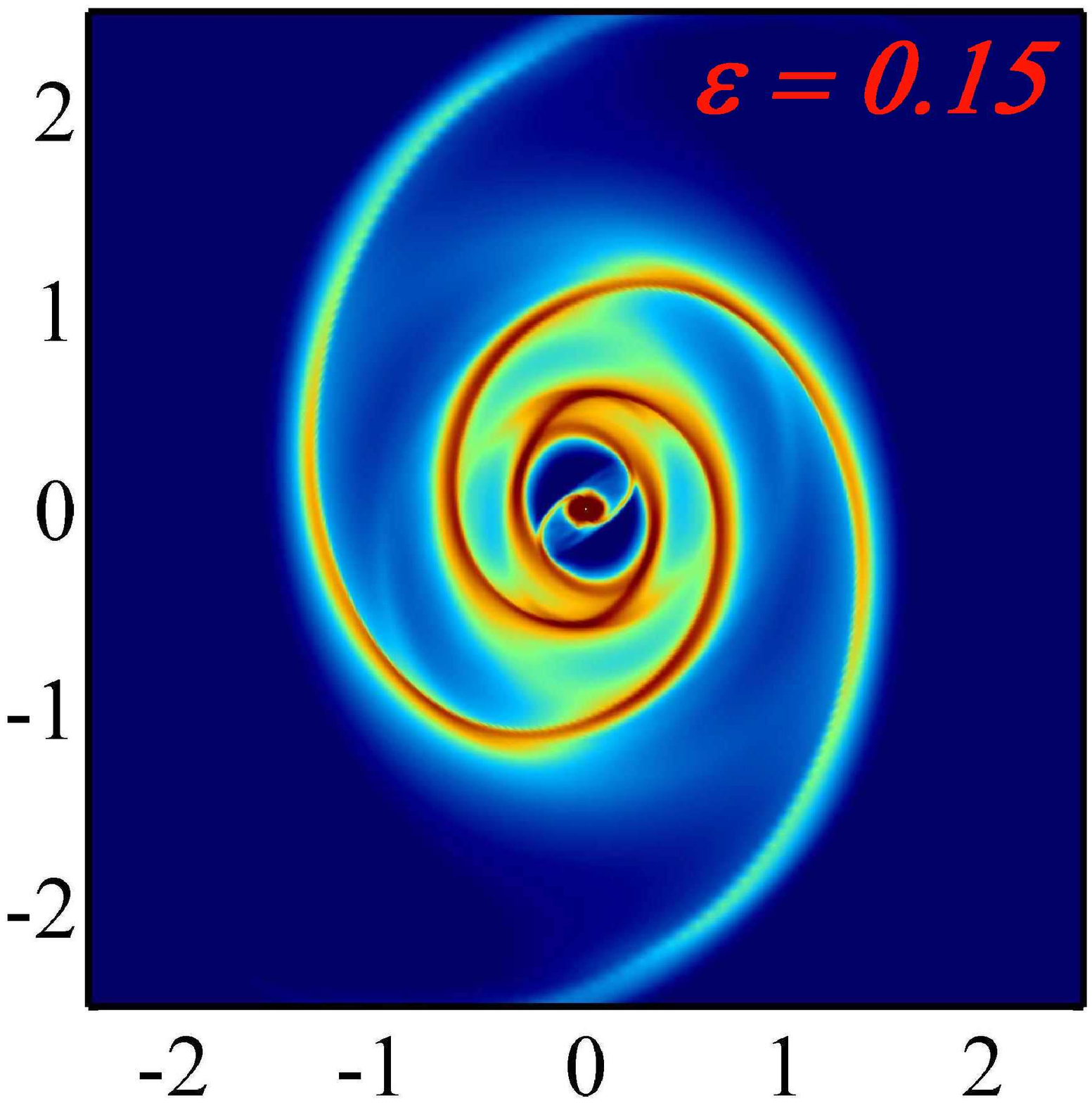}
 }

\vskip  0.0\hsize \hskip 0.0\hsize
  \vbox{\hsize=0.999\hsize
  \caption { Radial distributions of surface density along azimuth angle $\varphi=0^\circ$ for models with
  different~$\varepsilon$.}\label{Fig-profile-gas} }\vskip 0.0\hsize
\end{figure*}

Radial profiles of surface density are drawn on fig.~\ref{Fig-profile-gas} for a fixed azimuthal angle. The gas
disk responds even to 1 percent of halo non-axisymmetry. We see a small amplitude here. But for 5 percents of non-axisymmetry and
more shock waves appear.

Effects of gas temperature on the gas spiral pattern shown in
the fig.~\ref{Fig-mach}. Different values of the Mach number are
used. The circular rotation curves is fixed in these models. The
pitch angle of spirals decreases in more cold disk.

Kinematics of spiral pattern in the gas disk is similar to the
kinematics of the stellar disk submerged into nonaxisymmetric halo. We observe
characteristic quasiperiodic curves like in case of stellar disk.

Effect of selfgravitation increases the efficiency of spiral structure formation by the nonaxisymmetric halo. We simulate only gravitational stable disks. Morphology of waves changing and its amplitude grows.

\begin{figure*}
 \vskip 0.\hsize \hskip 0.0\hsize
 \centerline
{\includegraphics[width=0.5\hsize]%{empty.eps}%
{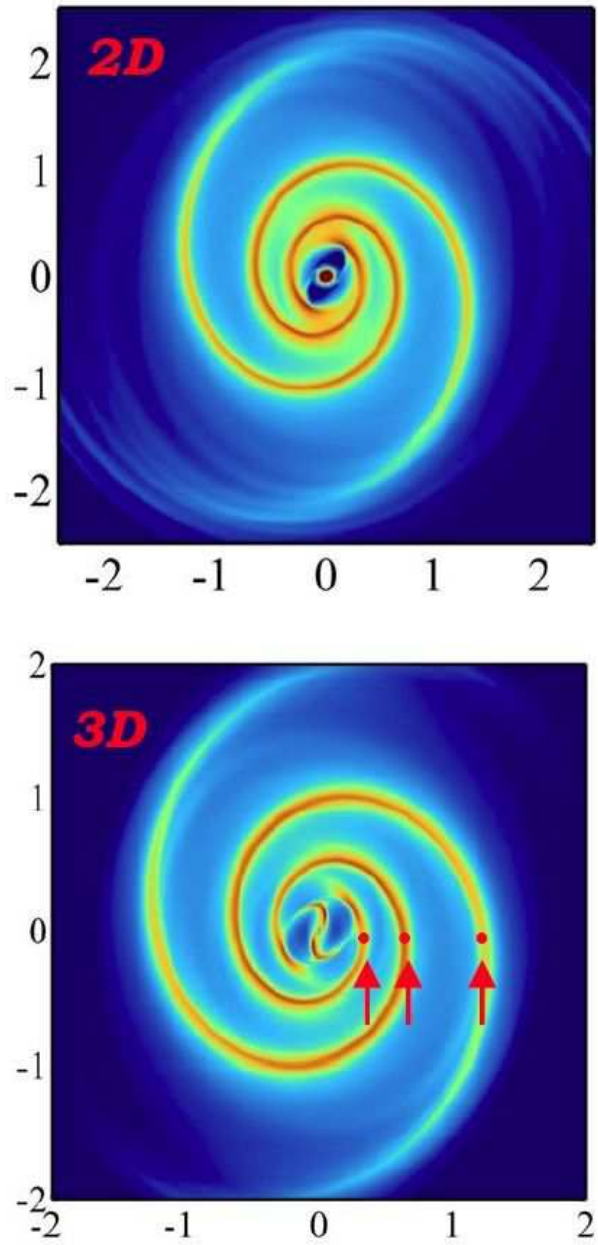}
} \vskip
0.0\hsize \hskip 0.0\hsize
  \vbox{\hsize=0.999\hsize
  \caption { It is a comparison of spiral patterns in 2D and 3D models.
  }\label{Fig-2d-3d-compare} }\vskip 0.0\hsize
\end{figure*}

\begin{figure*}
 \vskip 0.\hsize \hskip 0.0\hsize
\centerline
{\includegraphics[width=0.5\hsize]%{empty.eps}%
{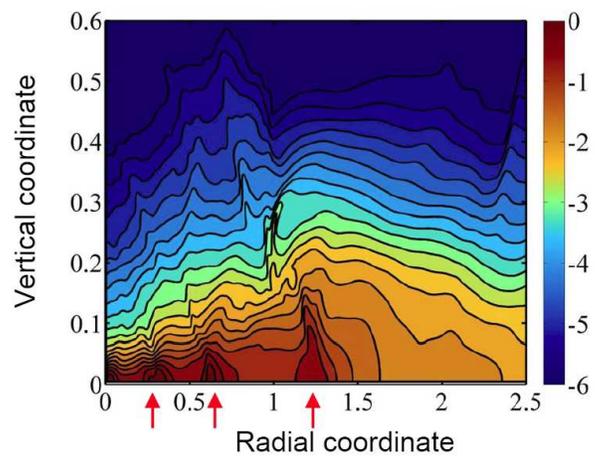}
} \vskip
0.0\hsize \hskip 0.0\hsize
  \vbox{\hsize=0.999\hsize
  \caption {  The vertical structure of gaseous disk submerged into triaxial dark halo. }\label{Fig-3d-vertical} }\vskip 0.0\hsize
\end{figure*}

We also demonstrate formation of similar spiral structures in 2d and 3d numerical simulations (fig.\ref{Fig-2d-3d-compare}).
Vertical distribution of volume density demonstrate vertical structure of density waves along radius (fig.\ref{Fig-3d-vertical}). Density waves marked by red arrows and dots at the disk plane (fig.\ref{Fig-2d-3d-compare} down) and also in vertical distribution at the same moment.

\begin{center}
  \textbf{5. Discussion and Conclusions}
\end{center}

\begin{enumerate}
    \item Possibility of 2-arms spiral structure formation in the stellar and gas components was shown in case of nonaxisymmetric dark massive
    halo. Properties of spiral pattern depends on parameters of disk and dark halo.

    \item Spiral patterns  changes in time.

    \item In general the spiral patterns have a similar morphology for 2D and 3D models.

    \item The spiral pattern rotates slowly. Position of corotation radius is at the outer part of the disk.

    \item A possibility to apply the mechanism consider for the formation of inner spiral structure remains open.
    It is also possible that the non-axisymmetric halo may explain the outer spiral patterns often observed in HI line and/or in
    UV (GALEX).

\end{enumerate}

We would like to thank A.Zasov and V.Korchagin for their useful
discussions. Numerical computations were carried out on CKIF--MGU
``Chebyshov'' at the Moscow State University, Moscow, Russia. This
work has been supported by RFBR ${\cal N}$09-02-97021.

\vfill
\newpage
\vfill

\centerline\textbf{ References }

\begin{enumerate}

  \item\noindent\label{Allgood-etal-2006}
Allgood B. et al. MN, 2006, \textbf{367}, 1781

  \item\noindent\label{Bailin-Steinmetz-2005}
Bailin J. \& Steinmetz M. ApJ, 2005, \textbf{627}, 647

  \item\noindent\label{Berman-2003!Floc-simul}
  Berman S.L. A\&A, 2003, \textbf{412}, 387

 \item\noindent\label{Bottema-2003!Simulat}
 Bottema R. MN, 2003, \textbf{344}, 358

\item\noindent\label{Bottema-Gerritsen-1997!NGC6503-model} Bottema
R., Gerritsen J.P.E. MN, 1997, \textbf{290}, 585

  \item\noindent\label{Brown-etal-2008}
Brown G.E., Geller D., Kenyon S. ApJ, 2008, \textbf{680}, 312

  \item\noindent\label{Bullock-2002}
Bullock J.S.  The Shapes of Galaxies and their Dark Halos. 2002,
109

   \item\noindent\label{Fridman-2008!UFN-instab}
Fridman A.M.  Phys. Usp., 2008, \textbf{51}, 213

    \item\noindent\label{Monaghan-sph}
    Monaghan J.J., Smoothed Particle Hydrodynamics, A\&A, 1992, \textbf{30}, 543

   \item\noindent\label{Iodice-etal-2003}
Iodice E., Arnaboldi M., Bournaud F. et al.  ApJ, 2003,
\textbf{585}, 730

  \item\noindent\label{Jing-Suto-2005}
Lee J., Jing Y.P., Suto Ya. ApJ, 2005, \textbf{632}, 706

  \item\noindent\label{Johnston-etal-2005}
Johnston K.V., Law D.R., Majewski S.R. ApJ, 2005, \textbf{619},
800

   \item\noindent\label{Helmi-2004}
Helmi A., Navarro J.F., Meza A., Steinmetz M., Eke V.R.
 ASPC, 2004, \textbf{327}, 87

    \item\noindent\label{Gnedin-etal-2005}
Gnedin O.Y., Gould A., Miralda-Escude J., Zentner A.R. ApJ, 2005,
\textbf{634}, 344

     \item\noindent\label{Gualandris-Zwart-2007}
Gualandris A., Zwart P., MN, 2007, \textbf{376}, 29

     \item\noindent\label{Kalberla-etal-2007}
Kalberla P.M.W., Dedes L., Kerp J., Haud U.
  A\&A, 2007, \textbf{469}, 511

  \item\noindent\label{Keller-etal-2008}
 Keller S., Murphy S. et al., ApJ, 2008, \textbf{678}, 851

 \item\noindent\label{Khoperskov-2002!Late-type}
 Khoperskov A.V. Astr. Letters, 2002, \textbf{28}, 651

 \item\noindent\label{Khoperskov-etal-2003}
 Khoperskov A.V., Zasov A.V., Tyurina N.V. Astr. Reports, 2003, \textbf{47}, 357

 \item\noindent\label{Khoperskov-Tyurina-2003!MW}
Khoperskov A.V., Tyurina N.V. Astr. Reports, 2003, \textbf{47},
443

 \item\noindent\label{Khoperskov-etal-2010!z-str}
 Khoperskov A.V.,  Bizyaev D., Tyurina N.V.,  Butenko M.  Astron. Nachr., 2010, \textbf{331}, 731

 \item\noindent\label{Korchagin-etal-2000!N1566}
 Korchagin V., Kikuchi N.,  Miyama S.M., Orlova N.,  Peterson B.A.
ApJ, \textbf{541}, 565

 \item\noindent\label{Kuhlen-etal-2007}
Kuhlen M., Diemand J., Madau P. ApJ, 2007, \textbf{671}, 1135

  \item\noindent\label{Law-Majewski-2010}
Law D., Majewski S., 2010arXiv1005.5390

  \item\noindent\label{Mikhailova-etal-2001!z-str}
Mikhailova E.A., Khoperskov A.V., Sharpak S.S. Stellar Dynamics:
from Classic to Modern (Eds. L.P. Osipkov and I.I. Nikiforov), St.
Petrsburg: St. Petersburg State Univ. Press, 2001, p.147

  \item\noindent\label{Brosch-Moiseev-2010}
 Brosch N., Kniazev A., Moiseev A., Pustilnik S.A. MN, 2010, \textbf{401}, 2067

  \item\noindent\label{Jeon-etal-2009!Galactic-Warps-Triaxial-Halos}
Jeon M., Kim S.S., Ann H.B. Galactic Warps in Triaxial Halos. ApJ,
2009, \textbf{696}, 1899

  \item\noindent\label{Narayan-etal-2005}
Narayan C.A., Saha K., Jog C.J. A\&A, 2005, \textbf{440}, 523

  \item\noindent\label{Navarro-etal-1997}
 Navarro J.F., Frenk C.S., White S. ApJ, 1997, \textbf{490}, 493

 \item\noindent\label{Newberg-Yanny-2005}
 Newberg H.J. \& Yanny B. ASPC, 2005, \textbf{338}, 210

 \item\noindent\label{Newberg-Yanny-2006}
 Newberg H.J. \& Yanny B., JPhCS, 2006, \textbf{47}, 195

  \item\noindent\label{Newberg-etal-2007}
Newberg H.J., Yanny B., Cole N., ApJ, 2007, \textbf{668}, 221

  \item\noindent\label{O'Brien-etal-2010}
O'Brien J.C., Freeman K.C., van der Kruit P.C. A\&A, 2010,
\textbf{515}, 63O

  \item\noindent\label{Olling-Merrifield-2000}
Olling R., Merrifield M. MN, 2000, \textbf{311}, 361

   \item\noindent\label{Patsis-etal-2009!dynamics-barred-spiral}
  Patsis P.A.,  Kaufmann D.E.,  Gottesman S.T.,  Boonyasait V. MN, 2009, \textbf{394}, 142

  \item\noindent\label{Reshetnikov-Sotnikova-2000}
 Reshetnikov V.P., Sotnikova N.Ya., Astr. Letter, 2000, \textbf{26}, 277

  \item\noindent\label{Roskar-etal-2010}
   Roskar R., Debattista V.P. et al. MN (1006.1659)

\item\noindent\label{Tutukov-Fedorova-2006}
 Tutukov A.V., Fedorova A.V.  Astr. Report, 2006, \textbf{50}, 785

 \item\noindent\label{Combes-2006}
Combes F., EAS, 2006, \textbf{20}, 97

 \item\noindent\label{Xu-etal-2006}
 Xu Y.,  Deng L.C.,  Hu J.Y. MN, 2006, \textbf{368}, 1811

 \item\noindent\label{Valluri-etal-2010}
 Valluri M., Debattista V.P. et al. MN, 2010, \textbf{403}, 525

 \item\noindent\label{Whitemore-etal-1987}
 Whitemore B.C., McElroy D.B., Schweizer F. ApJ, 1987, \textbf{314}, 439

 \item\noindent\label{Zasov-etal-2004}
 Zasov A.V., Khoperskov A.V., Tyurina N.V. Astr. Letters, 2004, \textbf{30}, 593

\end{enumerate}

%ÐÈÑðèñÐÈÑðèñÐÈÑðèñÐÈÑðèñÐÈÑðèñÐÈÑðèñÐÈÑðèñÐÈÑðèñÐÈÑðèñÐÈÑðèñ
%ÐÈÑðèñÐÈÑðèñÐÈÑðèñÐÈÑðèñÐÈÑðèñÐÈÑðèñÐÈÑðèñÐÈÑðèñÐÈÑðèñÐÈÑðèñ

%ÐÈÑðèñÐÈÑðèñÐÈÑðèñÐÈÑðèñÐÈÑðèñÐÈÑðèñÐÈÑðèñÐÈÑðèñÐÈÑðèñÐÈÑðèñ
%ÐÈÑðèñÐÈÑðèñÐÈÑðèñÐÈÑðèñÐÈÑðèñÐÈÑðèñÐÈÑðèñÐÈÑðèñÐÈÑðèñÐÈÑðèñ

%ÐÈÑðèñÐÈÑðèñÐÈÑðèñÐÈÑðèñÐÈÑðèñÐÈÑðèñÐÈÑðèñÐÈÑðèñÐÈÑðèñÐÈÑðèñ
%ÐÈÑðèñÐÈÑðèñÐÈÑðèñÐÈÑðèñÐÈÑðèñÐÈÑðèñÐÈÑðèñÐÈÑðèñÐÈÑðèñÐÈÑðèñ

%ÐÈÑðèñÐÈÑðèñÐÈÑðèñÐÈÑðèñÐÈÑðèñÐÈÑðèñÐÈÑðèñÐÈÑðèñÐÈÑðèñÐÈÑðèñ
%ÐÈÑðèñÐÈÑðèñÐÈÑðèñÐÈÑðèñÐÈÑðèñÐÈÑðèñÐÈÑðèñÐÈÑðèñÐÈÑðèñÐÈÑðèñ

%ÐÈÑðèñÐÈÑðèñÐÈÑðèñÐÈÑðèñÐÈÑðèñÐÈÑðèñÐÈÑðèñÐÈÑðèñÐÈÑðèñÐÈÑðèñ
%ÐÈÑðèñÐÈÑðèñÐÈÑðèñÐÈÑðèñÐÈÑðèñÐÈÑðèñÐÈÑðèñÐÈÑðèñÐÈÑðèñÐÈÑðèñ

%ÐÈÑðèñÐÈÑðèñÐÈÑðèñÐÈÑðèñÐÈÑðèñÐÈÑðèñÐÈÑðèñÐÈÑðèñÐÈÑðèñÐÈÑðèñ
%ÐÈÑðèñÐÈÑðèñÐÈÑðèñÐÈÑðèñÐÈÑðèñÐÈÑðèñÐÈÑðèñÐÈÑðèñÐÈÑðèñÐÈÑðèñ

%ÐÈÑðèñÐÈÑðèñÐÈÑðèñÐÈÑðèñÐÈÑðèñÐÈÑðèñÐÈÑðèñÐÈÑðèñÐÈÑðèñÐÈÑðèñ
%ÐÈÑðèñÐÈÑðèñÐÈÑðèñÐÈÑðèñÐÈÑðèñÐÈÑðèñÐÈÑðèñÐÈÑðèñÐÈÑðèñÐÈÑðèñ

%ÐÈÑðèñÐÈÑðèñÐÈÑðèñÐÈÑðèñÐÈÑðèñÐÈÑðèñÐÈÑðèñÐÈÑðèñÐÈÑðèñÐÈÑðèñ
%ÐÈÑðèñÐÈÑðèñÐÈÑðèñÐÈÑðèñÐÈÑðèñÐÈÑðèñÐÈÑðèñÐÈÑðèñÐÈÑðèñÐÈÑðèñ

%ÐÈÑðèñÐÈÑðèñÐÈÑðèñÐÈÑðèñÐÈÑðèñÐÈÑðèñÐÈÑðèñÐÈÑðèñÐÈÑðèñÐÈÑðèñ
%ÐÈÑðèñÐÈÑðèñÐÈÑðèñÐÈÑðèñÐÈÑðèñÐÈÑðèñÐÈÑðèñÐÈÑðèñÐÈÑðèñÐÈÑðèñ

%ÐÈÑðèñÐÈÑðèñÐÈÑðèñÐÈÑðèñÐÈÑðèñÐÈÑðèñÐÈÑðèñÐÈÑðèñÐÈÑðèñÐÈÑðèñ
%ÐÈÑðèñÐÈÑðèñÐÈÑðèñÐÈÑðèñÐÈÑðèñÐÈÑðèñÐÈÑðèñÐÈÑðèñÐÈÑðèñÐÈÑðèñ

%ÐÈÑðèñÐÈÑðèñÐÈÑðèñÐÈÑðèñÐÈÑðèñÐÈÑðèñÐÈÑðèñÐÈÑðèñÐÈÑðèñÐÈÑðèñ
%ÐÈÑðèñÐÈÑðèñÐÈÑðèñÐÈÑðèñÐÈÑðèñÐÈÑðèñÐÈÑðèñÐÈÑðèñÐÈÑðèñÐÈÑðèñ

%ÐÈÑðèñÐÈÑðèñÐÈÑðèñÐÈÑðèñÐÈÑðèñÐÈÑðèñÐÈÑðèñÐÈÑðèñÐÈÑðèñÐÈÑðèñ
%ÐÈÑðèñÐÈÑðèñÐÈÑðèñÐÈÑðèñÐÈÑðèñÐÈÑðèñÐÈÑðèñÐÈÑðèñÐÈÑðèñÐÈÑðèñ

\end{document}